\documentclass[pra,twocolumn,longbibliography]{revtex4-2}
\usepackage{graphicx}  
\usepackage{dcolumn}   
\usepackage{bm}        
\usepackage{verbatim}   
\usepackage{graphicx}
\usepackage{epstopdf}
\usepackage{footnote}
\usepackage{epsfig}
\usepackage{subfigure}
\usepackage{amssymb}
\usepackage{amsmath,bm}
\usepackage{xcolor}
\usepackage{float}
\usepackage{subfigure}
\usepackage[most]{tcolorbox}
\usepackage{soul}
\usepackage{color}
\usepackage[colorlinks=true,linkcolor=blue,allcolors=blue]{hyperref}%
\usepackage[framemethod=TikZ]{mdframed}
\usepackage{lipsum}
\mdfdefinestyle{MyFrame}{%
	linecolor=blue,
	outerlinewidth=2pt,
	roundcorner=20pt,
	innertopmargin=\baselineskip,
	innerbottommargin=\baselineskip,
	innerrightmargin=20pt,
	innerleftmargin=20pt,
	backgroundcolor=gray!50!white}

\begin{document}
	
	
\headsep = 40pt
\title{Enhancement of the Environmental Stability of Perovskite Thin Films via PMMA and AZ5214-Photoresist Coatings}
\author{Kimya Fallah$^1$, Shahab Norouzian Alam$^{1,2}$, Bijan Ghafary Ghomi$^1$, Farzaneh Yekekar$^1$, Shima Taghian$^1$, and Sajjad Taravati$^3$}

\affiliation{$^1$Physics Department, Iran University of Science and Technology, Tehran, Iran\\
	$^2$Optoelectronic Research Center, Iran University of Science and Technology\\
	$^3$School of Electronics and Computer Science, University of Southampton, Southampton SO171BJ, UK}
	
\begin{abstract}
We introduce a pioneering strategy to enhance the environmental stability of perovskite thin films, a critical step forward in advancing their application in optoelectronics. Through the innovative application of matrix encapsulation techniques, we focus on the stabilization of methylammonium lead iodide (MAPbI$_3$) and methylammonium lead bromide (MAPbBr$_3$) films. These films are meticulously prepared via a two-step solution deposition method under controlled ambient conditions. Our approach involves the application of poly(methyl methacrylate) (PMMA) and AZ5214 photoresist layers through spin-coating, aimed at singularly encapsulating the perovskite films. This encapsulation acts as a robust hydrophobic barrier, significantly mitigating moisture ingress and simultaneously addressing the challenge of pinhole presence within the perovskite structure. Through a series of detailed characterizations—spanning scanning electron microscopy (SEM), X-ray diffraction (XRD), and photoluminescence (PL) spectroscopy—we demonstrate that, despite the thicker nature of the AZ5214 photoresist compared to the PMMA layer, it exhibits markedly enhanced stability. Notably, the integrity and optical properties of the perovskite films are preserved for extended periods of up to 960 hours under environmental exposure. This breakthrough highlights the superior performance of AZ5214 photoresist over PMMA in prolonging the operational life of perovskite thin films, thereby offering a promising avenue for their deployment in a wide range of optoelectronic devices.
\end{abstract}
	
\maketitle

\section{Introduction}

Lead halide perovskites (LHPs) have risen to prominence as exceptional optoelectronic materials, thanks to their superior photovoltaic power conversion efficiencies~\cite{PhysRevLett.132.086902,PhysRevApplied.13.014005} and cost-effective production~\cite{PhysRevLett.121.085502}. They exhibit a crystalline structure similar to calcium titanium oxide (CaTiO$_3$), adopting the general formula ABX$_3$. In this formula, "A" represents a cation, such as methylammonium (CH$_3$NH$_3^+$), formamidinium (CH(NH$_2$)$_2^+$), or cesium; "B" denotes lead (Pb); and "X" is a halide anion, which can be chloride (Cl$^-$), bromide (Br$^-$), or iodide (I$^-$). These compounds are part of the wider perovskite material family, distinguished by their unique and adaptable crystalline framework. They boast impressive features, including long charge carrier diffusion lengths~\cite{de2015impact}, precise tunable bandgaps~\cite{jeon2018fluorene}, high light absorption coefficients~\cite{brenner2016hybrid,qin2014inorganic,PhysRevB.101.054108}, electronic correlations and unusual excitonic effects~\cite{PhysRevX.8.021034}, and exceptional defect tolerance~\cite{akkerman2018genesis,PhysRevMaterials.6.055402}. The exceptional photovoltaic characteristics of these optoelectronic materials, as detailed in existing literature~\cite{stranks2015metal,PhysRevLett.132.086902}, position them as a highly promising foundation for the development of next-generation, high-frequency, dynamic metasurfaces~\cite{PhysRevApplied.12.024026,PhysRevApplied.14.014027,Taravati_NC_2021,Taravati_ACSP_2022}. In addition, LHPs are excellent candidates for creation of exciton polariton at room temperature thanks to their large exciton binding energy and quantum yield~\cite{PhysRevApplied.18.014079}. Such attributes make LHPs particularly well-suited for the fabrication of optoelectronic devices~\cite{ye2016soft,PhysRevApplied.10.041001}, solar cells~\cite{green2014emergence,PhysRevApplied.19.054039,qin2014inorganic}, lasers~\cite{jiang2018continuous}, light-emitting diodes (LEDs)~\cite{stranks2015metal,xing2018color}, quantum confinement and dielectric deconfinement~\cite{PhysRevApplied.17.054045}, photovoltaic and photocatalysis applications~\cite{PhysRevApplied.21.014063}, and photodetectors~\cite{saidaminov2015planar}.

However, despite their immense potential, LHP materials face significant challenges when exposed to external factors such as moisture and heat, largely due to their minimal formation energy (approximately 0.1-0.3 eV)~\cite{leijtens2015stability,berhe2016organometal,wang2016stability}. Enhancing the stability of perovskite materials is therefore crucial for extending the lifespan and performance of devices based on these materials. Various methods have been explored to fortify LHPs against water ingress~\cite{guarnera2015improving,huang2016enhancing,di2015size}, yet many of these approaches can lead to the generation of surface lead ions and the formation of easily exfoliated coatings due to the presence of fragile chemical bonds~\cite{tannenbaum2006ftir}. To address these challenges, researchers have investigated the incorporation of poly methyl methacrylate (PMMA) onto LHP nano-crystals (NCs)~\cite{tannenbaum2006ftir,tannenbaum2004infrared,ciprari2006characterization}. The ester carbonyl groups of PMMA contribute to the formation of high-quality composite LHP films, characterized by pure colors and exceptional resistance to both heat and water~\cite{li2018stable}.

In this study, we present an exploration into enhancing the stability of perovskite structures by applying nanostructured layers of PMMA and AZ5214 photoresist separately. We synthesize MAPbI$_3$ and MAPbBr$_3$ films with single-side ultra-thin layers of PMMA polymer and AZ5214 photoresist individually. Structural and optical properties are systematically investigated using X-ray diffraction (XRD), photoluminescence (PL), and scanning electron microscopy (SEM). Our findings reveal that the utilization of the AZ5214 photoresist layer effectively shields the perovskite layer against environmental factors, demonstrating comparable performance to the PMMA layer in this regard.

\section{Materials and methods}\label{sec:}  

\subsection{Materials}

Glass slides (1×1 cm), lead iodide powder (PbI$_2$), methylammonium iodide powder (MAI), PMMA powder, acetone, methylammonium bromide powder (MABr), lead bromide powder (PbBr$_2$), dimethyl sulfoxide (DMSO), AZ5214 photoresist, and chlorobenzene were sourced from Merk company and utilized as received without additional purification. 
Scanning electron microscope XL300 manufactured by Philips. X-ray diffraction spectrometer Expert manufactured by Philips. Photoluminescence device Noora 200.

\begin{figure}
	\begin{center}
		\includegraphics[width=1\columnwidth]{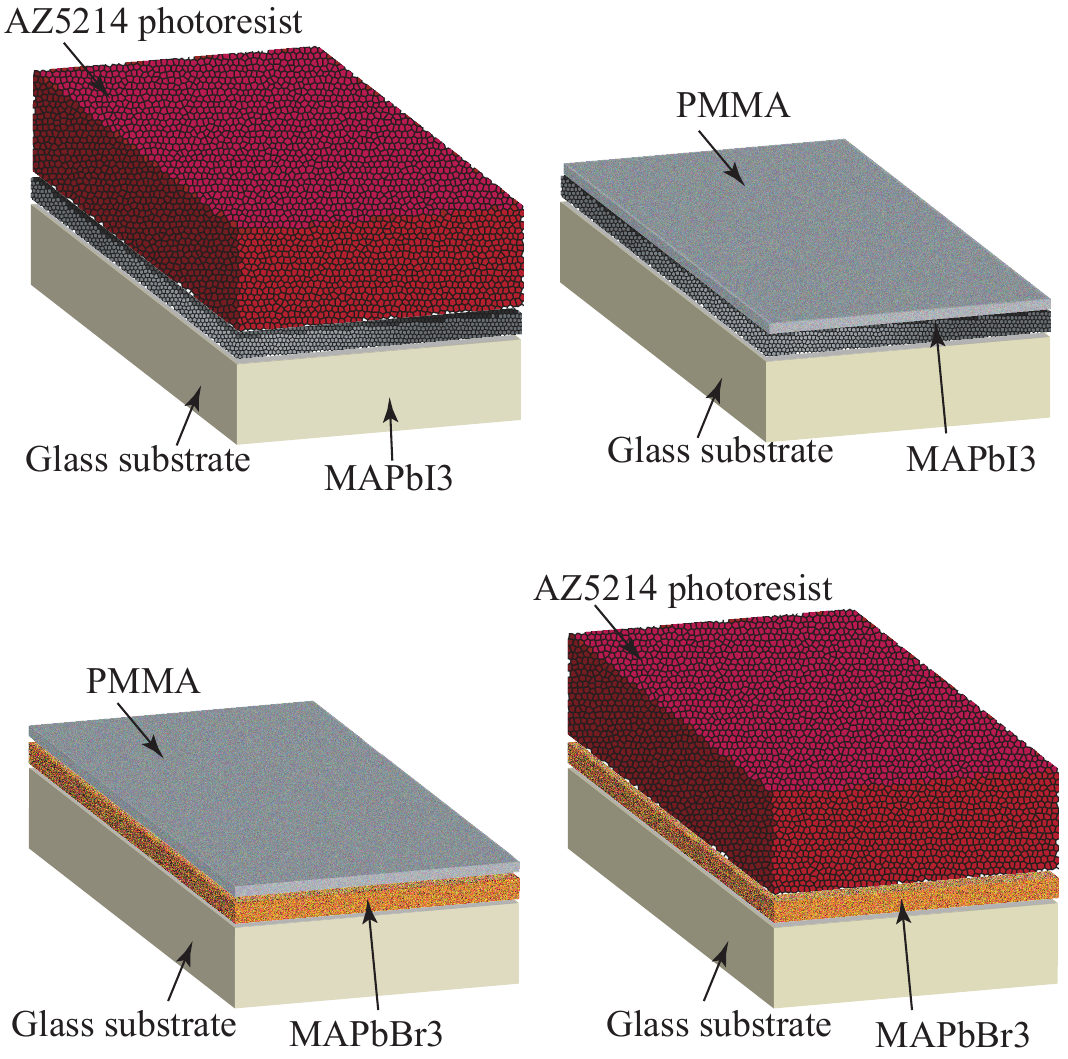}
		\caption{Enhancing the stability of MAPbI$_3$ and MAPbBr$_3$ perovskite structures by applying PMMA and AZ5214 photoresist coatings.}
		\label{fig:sch} 
	\end{center}
\end{figure}

\begin{figure}
	\begin{center}
		\includegraphics[width=1\columnwidth]{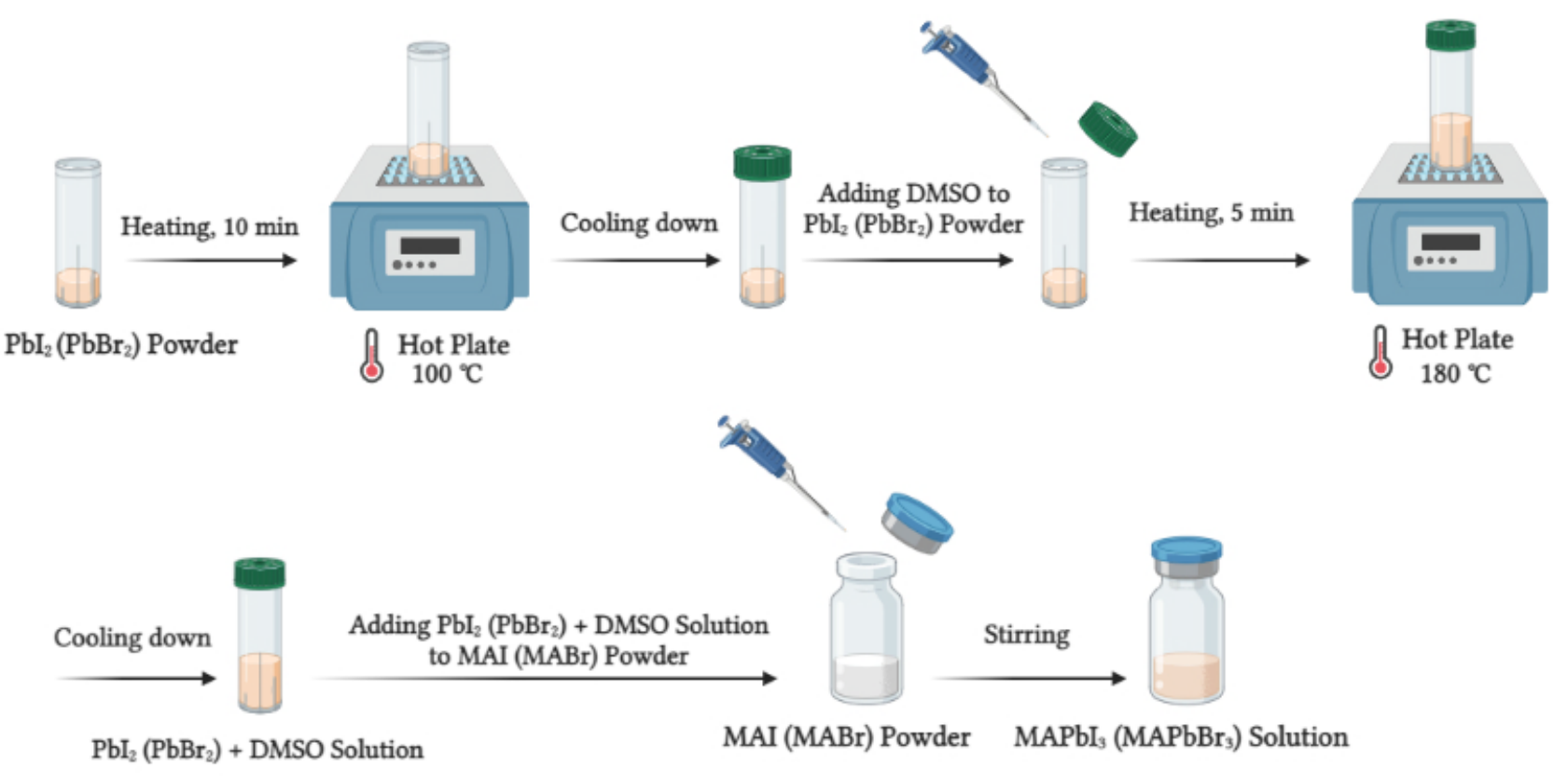}
		\caption{Synthesis of MAPbI$_3$ and MAPbBr$_3$ solutions.}
		\label{fig:1} 
	\end{center}
\end{figure}

\subsection{Synthesis of Perovskite Solution and PMMA Polymer}

Figure 1 demonstrates a schematic representation of the proposed solutions for enhancing the stability of MAPbI$_3$ and MAPbBr$_3$ perovskite structures by applying PMMA and AZ5214 photoresist coatings. The precursor solutions for MAPbI$_3$ and MAPbBr$_3$ were prepared by dissolving 55 mg of MAI (MABr) and 190 mg of PbI$_2$ (PbBr$_2$), respectively, with a 3:1 molar ratio in 300.5 $\mu l$ of DMSO at 180°C. To ensure dehumidification, the lid of the PbI$_2$ (PbBr$_2$) vial (190 mg for PbI$_2$ and 181 mg for PbBr$_2$) was opened and placed on a hotplate at 100°C for 10 minutes, sealed, and then allowed to cool to ambient temperature. Subsequently, 300.5 $\mu l$ of DMSO was added at 180°C and stirred until completely dissolved. The final precursor solution for MAPbI$_3$ (MAPbBr$_3$) was obtained by adding 55 mg of MAI (MABr) to 291 $\mu l$ of the above solution, followed by chilling and continuous stirring for 5 minutes, as depicted in Fig. 2. For the preparation of the PMMA polymer solution, 25 mg of PMMA powder was dissolved in 500 $\mu l$ of acetone by stirring for 30 minutes.

\subsection{Fabrication of Perovskite Thin Film}
The glass substrates underwent a rigorous cleaning process involving sequential immersion in water and soap, double-distilled water, and ethanol, followed by ultrasonication at 70°C for 10 minutes after each step. Subsequently, the substrates were dried using a heat gun and then placed in a laboratory furnace at 500°C for one hour to ensure thorough drying.

To fabricate MAPbI$_3$ (MAPbBr$_3$) thin films, the spin-coating technique, specifically the nanocrystal-pinning (NCP) process, was employed in two steps. The cleaned substrate was mounted on a spin-coater, and 25 $\mu l$ of MAPbI$_3$ (MAPbBr$_3$) precursor solution was dispensed onto it, followed by acceleration to 1000 rpm for 10 seconds and then to 4000 rpm for 30 seconds. At the 15-second mark of the second step, 100 $\mu l$ of chlorobenzene solution was dispensed onto the spinning substrate. Subsequent evaporation of the remaining solution facilitated rapid crystallization of the perovskite nanocrystals (NCs). For additional annealing, the substrate was placed on a hotplate at 100°C for 40 minutes.

Once the film had cooled, a layer of PMMA polymer (or AZ5214 photoresist) was applied atop the perovskite layer. The spin-coating process was repeated, with acceleration to 3000 rpm (6000 rpm for AZ5214 photoresist) for 30 seconds, followed by placement on a hotplate at 100°C for 10 minutes to ensure proper adhesion and drying. Representative samples are depicted in Figs. 3 and 4.

\begin{figure}
	\begin{center}
		\subfigure[]{\label{fig:2a} 
			\includegraphics[width=0.3\columnwidth]{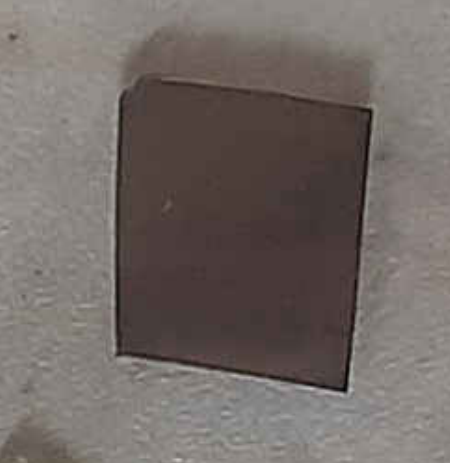}} 
		\subfigure[]{\label{fig:2b} 
			\includegraphics[width=0.3\columnwidth]{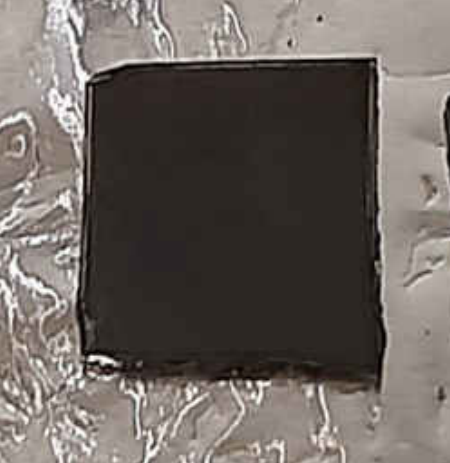}}
		\subfigure[]{\label{fig:2c} 
		\includegraphics[width=0.3\columnwidth]{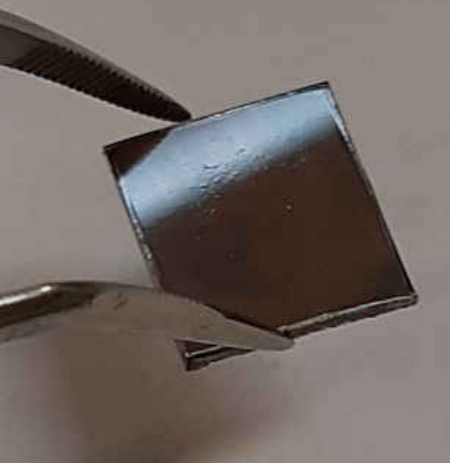}}
		\caption{Images of MAPbI$_3$ perovskite samples: (a) Perovskite, (b) Perovskite with PMMA, and (c) Perovskite with AZ5214 photoresist.} 
		\label{Fig:2}
	\end{center}
\end{figure}

\begin{figure}
	\begin{center}
		\subfigure[]{\label{fig:3a} 
			\includegraphics[width=0.3\columnwidth]{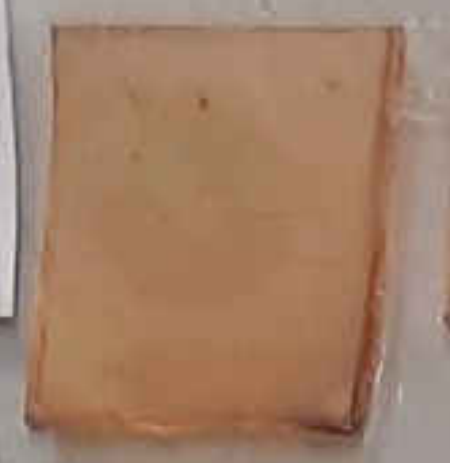}} 
		\subfigure[]{\label{fig:3b} 
			\includegraphics[width=0.3\columnwidth]{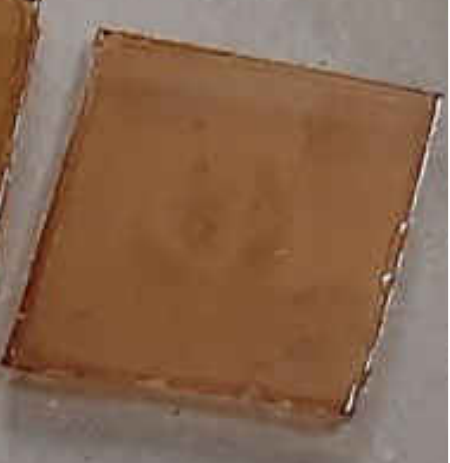}}
		\subfigure[]{\label{fig:3c} 
			\includegraphics[width=0.3\columnwidth]{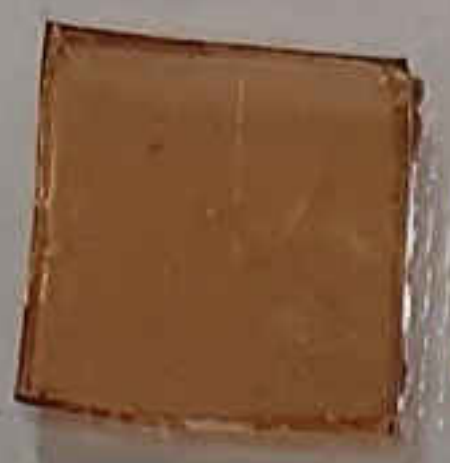}}
		\caption{Images of MAPbBr$_3$ perovskite samples: (a) Perovskite, (b) Perovskite with PMMA, and (c) Perovskite with AZ5214 photoresist.} 
		\label{Fig:3}
	\end{center}
\end{figure}

\section{Results}

\subsection{Morphological Analysis of MAPbI$_3$ Perovskite Thin Films using SEM}

The morphology of the MAPbI$_3$ perovskite thin layer was investigated through SEM, including top-view (SEM-Top) and cross-sectional (SEM-Cross). Figures 5(a) to 5(c) present SEM-Top analysis results for the optimal perovskite layer at three different resolutions, that is, 50 $\mu$m, 5 $\mu$m and 1 $\mu$m. Figures 6(a) and 6(b) provide a comparative analysis of the SEM-Top images of the perovskite surface, perovskite with PMMA, and perovskite with AZ5214 photoresist. The application of PMMA and AZ5214 photoresist does not alter the inherent structure of the perovskite; instead, it serves to encapsulate the perovskite material, creating a protective layer akin to plastic. This encapsulation process enhances the stability of the perovskite against various environmental factors.  

\begin{figure}
	\begin{center}
		\subfigure[]{\label{fig:4a} 
			\includegraphics[width=0.85\columnwidth]{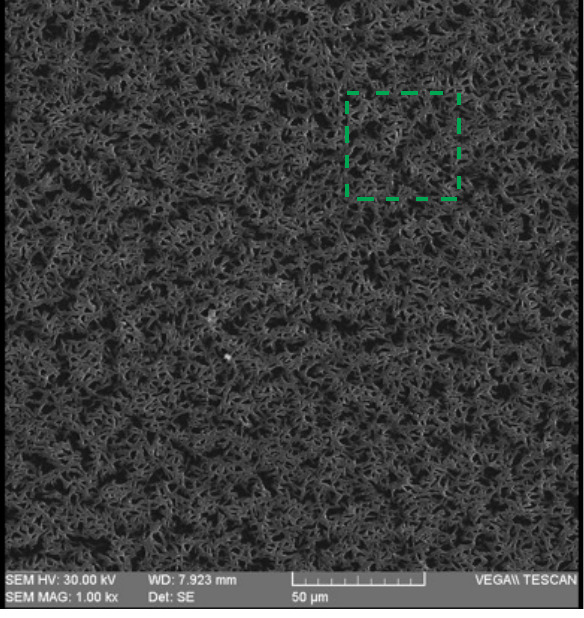}} 
		\subfigure[]{\label{fig:4b} 
			\includegraphics[width=0.48\columnwidth]{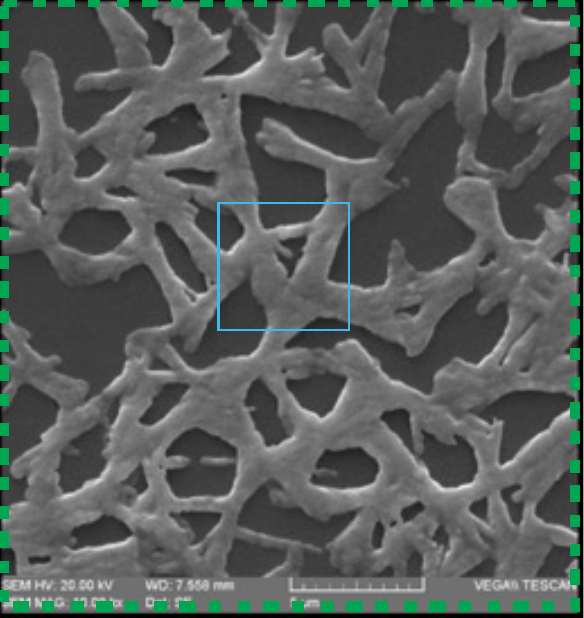}}
		\subfigure[]{\label{fig:4c} 
			\includegraphics[width=0.48\columnwidth]{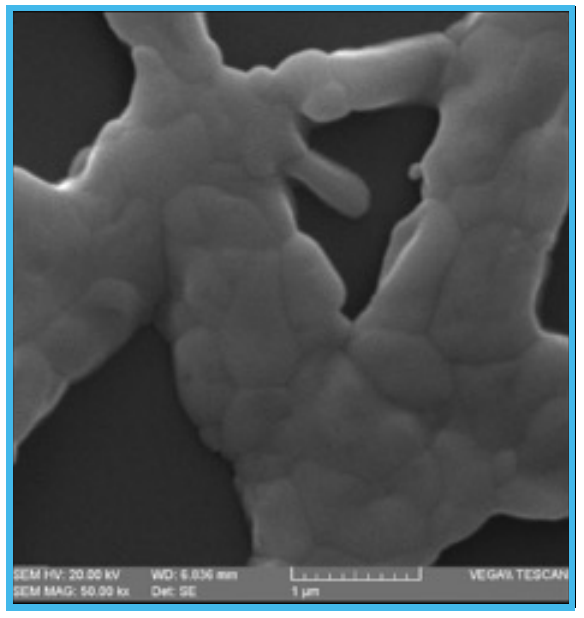}}
		\caption{SEM top images of the perovskite surface at three different resolutions. (a) 50 $\mu$m. (b) 5 $\mu$m. (c) 1 $\mu$m.} 
		\label{Fig:4}
	\end{center}
\end{figure}

\begin{figure}
	\begin{center}
		\subfigure[]{\label{fig:5a} 
			\includegraphics[width=0.85\columnwidth]{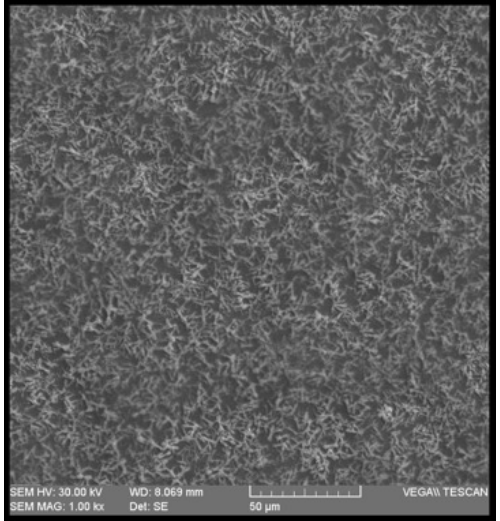}} 
		\subfigure[]{\label{fig:5b} 
			\includegraphics[width=0.85\columnwidth]{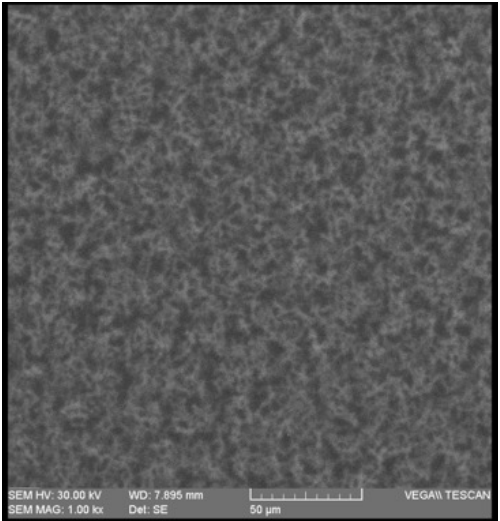}}
		\caption{SEM-Top images of the perovskite surface with (a) PMMA, and (b) AZ5214 photoresist, in comparison with the perovskite surface in Fig.~\ref{Fig:4}.} 
		\label{Fig:5}
	\end{center}
\end{figure}

Figures 7(a) to 7(c) show SEM-Cross analysis to check the thickness of perovskite, PMMA and AZ5214 photoresist layer. Initially, a perovskite layer with a thickness of 395 nanometers was deposited. Since the AZ5214 photoresist emits light in the range of 500 to 700 nanometers, the perovskite layer was further deposited to a thickness of 425 nanometers in order to partially shield the AZ5214 photoresist layer from this emission.

\begin{figure*}
	\begin{center}
		\subfigure[]{\label{fig:6a} 
			\includegraphics[width=0.65\columnwidth]{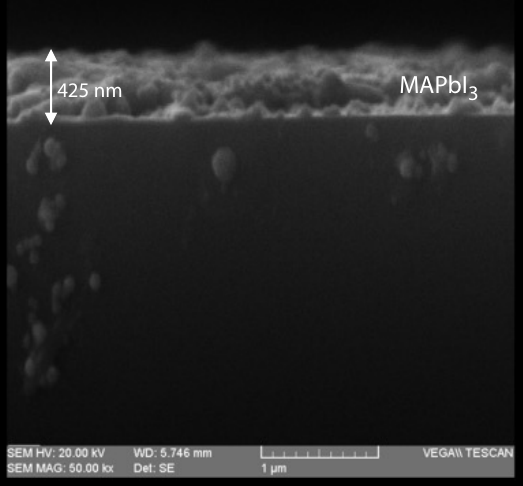}} 
		\subfigure[]{\label{fig:6b} 
			\includegraphics[width=0.65\columnwidth]{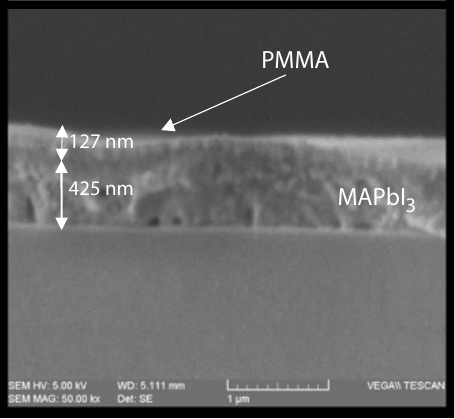}}
	     \subfigure[]{\label{fig:6c} 
				\includegraphics[width=0.65\columnwidth]{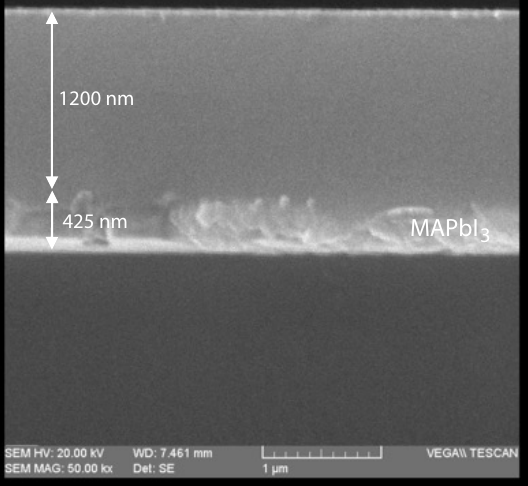}}
		\caption{Thickness comparison of (a) MAPbI$_3$ perovskite, (b) perovskite with PMMA, and (c) perovskite with AZ5214 photoresist.} 
		\label{Fig:6}
	\end{center}
\end{figure*}

\subsection{X-ray Diffraction (XRD) Analysis}

Figure 8 presents the X-ray diffraction (XRD) analysis of the perovskite phase of the MAPbI$_3$ film at two different time points: 24 hours and 40 days after deposition. This analysis aims to investigate the crystal structure and assess stability by comparing the obtained results with the XRD pattern from Ref.~\cite{wei2017importance}. In Figure 9, XRD analysis of the perovskite film, along with AZ5214 photoresist, one day and 40 days after deposition is presented. According to the obtained diagram, there is no change observed after 40 days. The average grain size ($D$) of the MAPbI$_3$ nanocrystals can be calculated based on the (002) peak observed at 14.2° using the Scherrer equation. This equation relates the grain size to the wavelength of X-rays ($\lambda$), the Bragg diffraction angle ($\theta$), the full width at half maximum (FWHM) of the (002) diffraction peak, and a constant factor ($k$). Specifically, the equation is expressed as $D = k \lambda/(\beta \cos(\theta))$, where $k$ is the Scherrer constant (typically 0.89) and $\lambda$ is the X-ray wavelength (0.154 nm). By applying this equation, one can accurately determine the average grain size of the nanocrystals. The FWHM of the (002) diffraction peak is 0.41, resulting in an average grain size of approximately 19 nanometers.

The consistent XRD patterns observed over a 40-day period underscore the structural integrity and stability of the MAPbI$_3$ perovskite films, even when integrated with AZ5214 photoresist. This stability, particularly in the absence of any significant structural changes over time, is indicative of a robust perovskite film. The calculated average grain size of 19 nanometers further validates the high-quality crystal structure of the MAPbI$_3$ nanocrystals within the film. The retention of crystal structure and grain size over an extended period highlights the effectiveness of our matrix encapsulation approach in preserving the structural attributes of the perovskite film. Such a stable and well-defined crystal structure is crucial for optoelectronic applications, as it directly impacts the material's optical and electronic properties. Therefore, the results presented here not only demonstrate the successful preparation of high-quality perovskite films but also affirm their potential longevity and performance in device applications. This alignment with the referenced XRD pattern from the literature further corroborates the high structural fidelity of our perovskite films, establishing a solid foundation for their application in stable and efficient optoelectronic devices.

\begin{figure}
	\begin{center}
		\includegraphics[width=1\columnwidth]{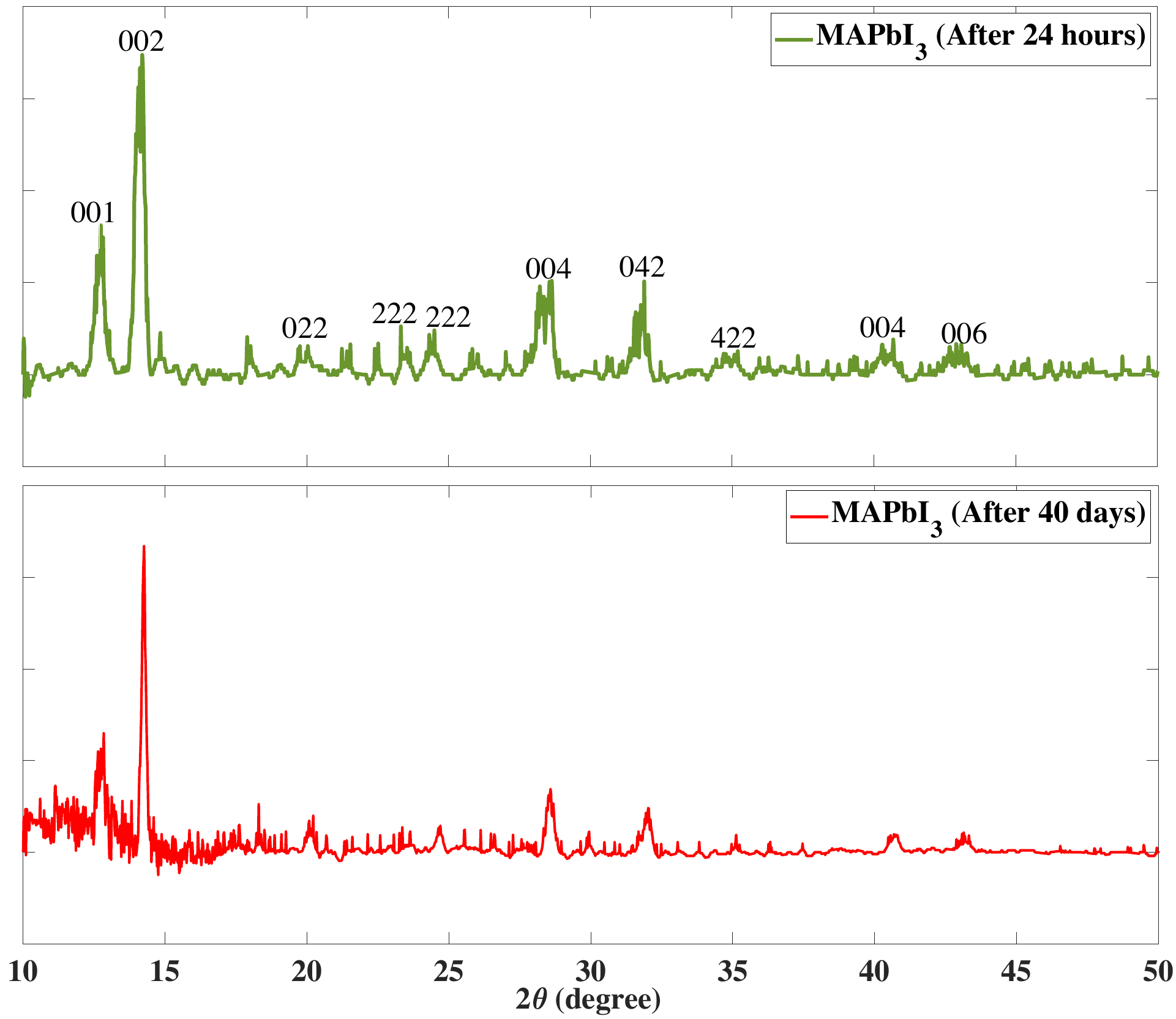}
		\caption{X-ray diffraction (XRD) analysis of the perovskite film 24 hours and 40 days after deposition.
			Angles of 12.74$^\circ$, 14.20$^\circ$, 19.76$^\circ$, 23.32$^\circ$, 24.50$^\circ$, 28.64$^\circ$, 31.90$^\circ$, 35.20$^\circ$, and 42.90$^\circ$ correspond to the (001), (002), (022), (222), (004), (042), (422), (044), and (006) planes, respectively. A decrease in peak intensity is observed after 40 days.}
		\label{fig:7} 
	\end{center}
\end{figure}
\begin{figure}
	\begin{center}
		\includegraphics[width=1\columnwidth]{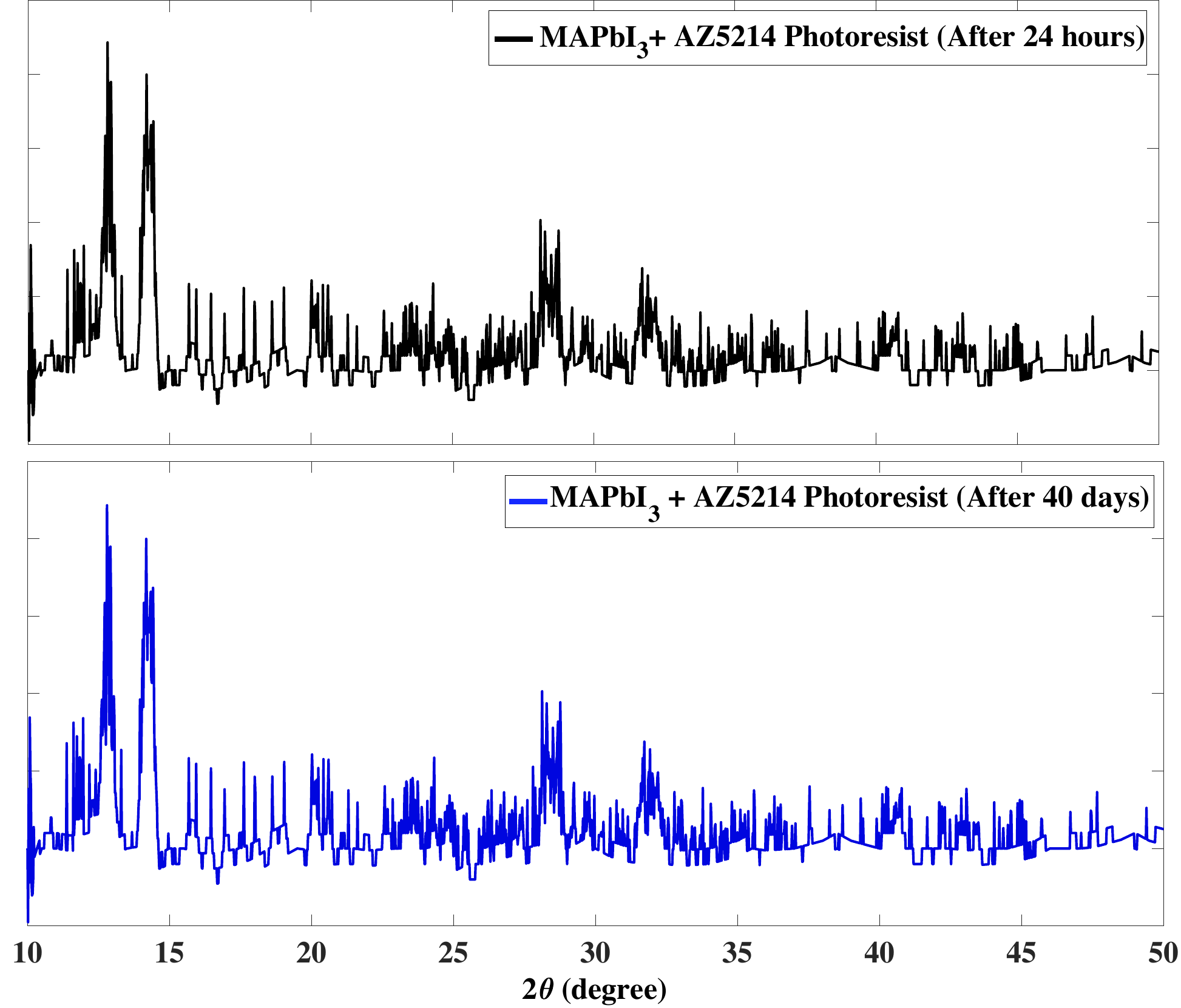}
		\caption{X-ray diffraction (XRD) analysis of the perovskite film along with AZ5214 photoresist 24 hours and 40 days after deposition.}
		\label{fig:8} 
	\end{center}
\end{figure}

\subsection{PL Spectra Analysis of MAPbI$_3$ and MAPbBr$_3$}

\begin{figure}
	\begin{center}
		\includegraphics[width=0.6\columnwidth]{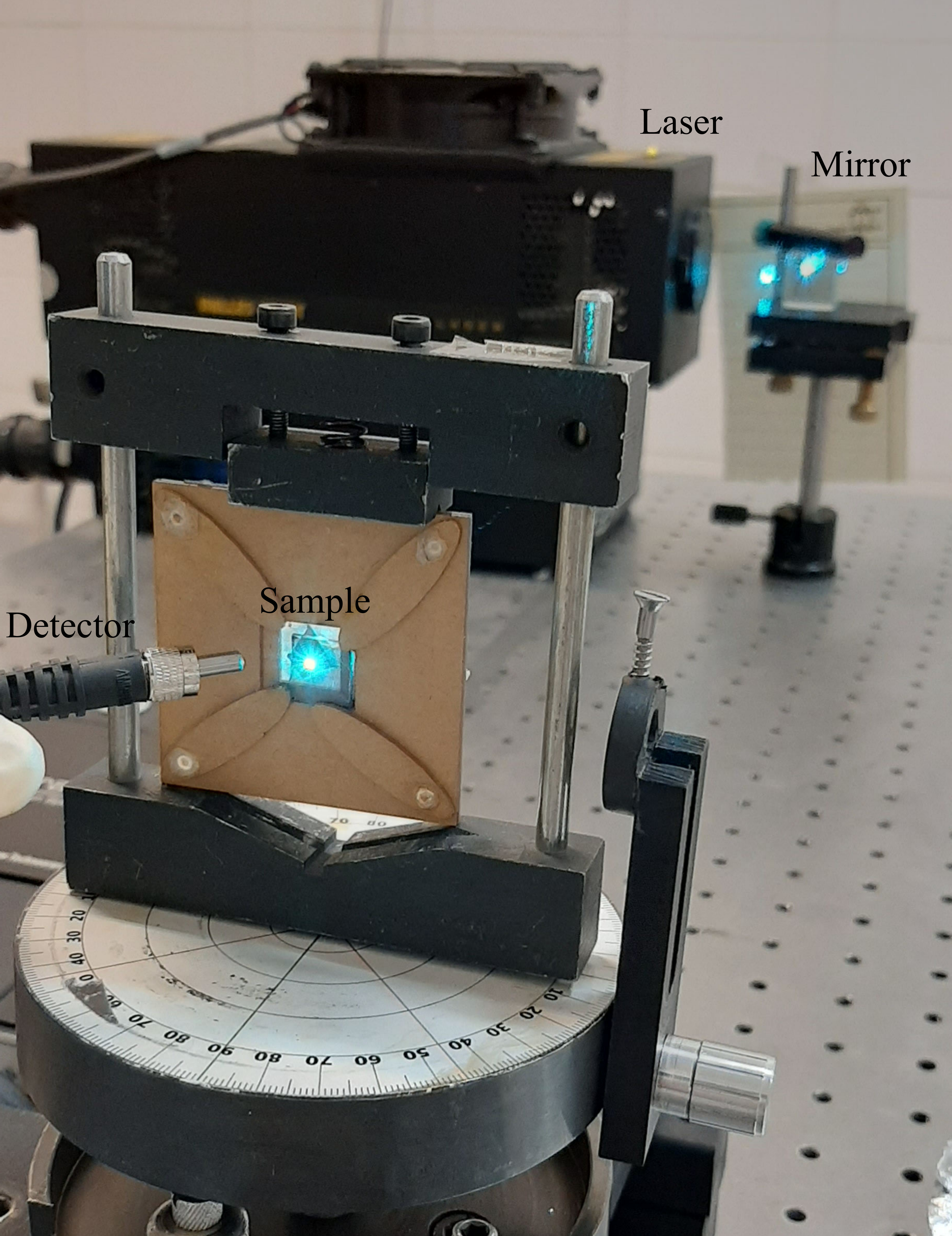}
		\caption{Experimental setup for photoluminescence (PL) spectroscopy.}
		\label{fig:8} 
	\end{center}
\end{figure}

\begin{figure}
	\begin{center}
		\subfigure[]{\label{fig:9a} 
			\includegraphics[width=0.47\columnwidth]{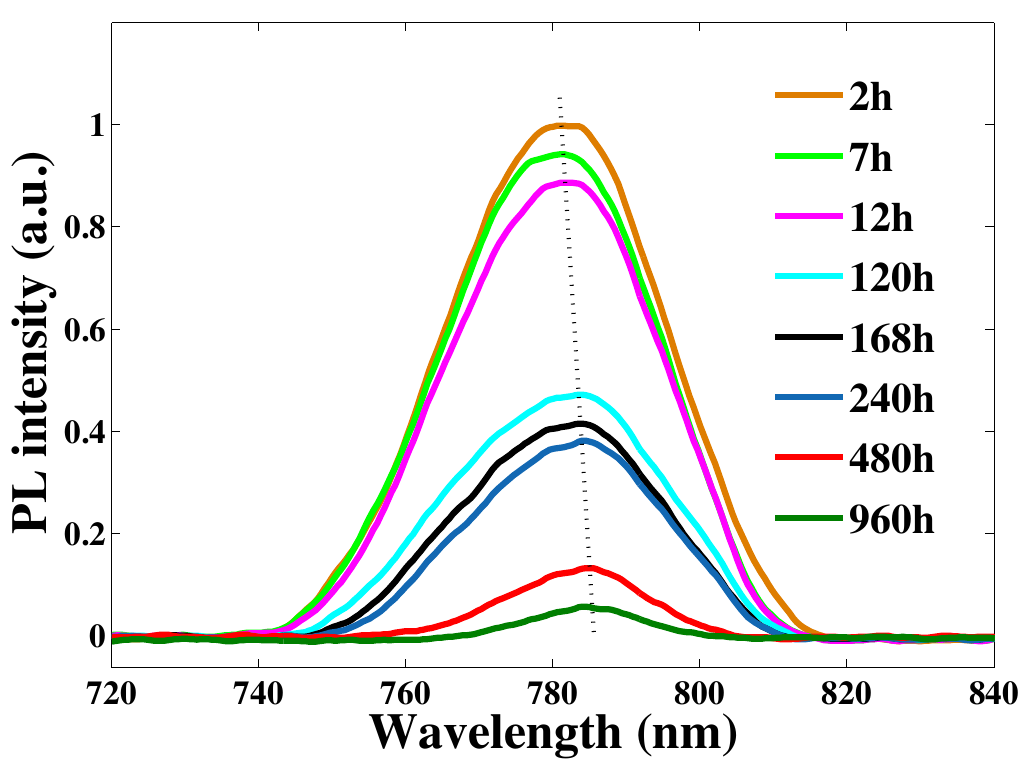}} 
		\subfigure[]{\label{fig:9b} 
			\includegraphics[width=0.48\columnwidth]{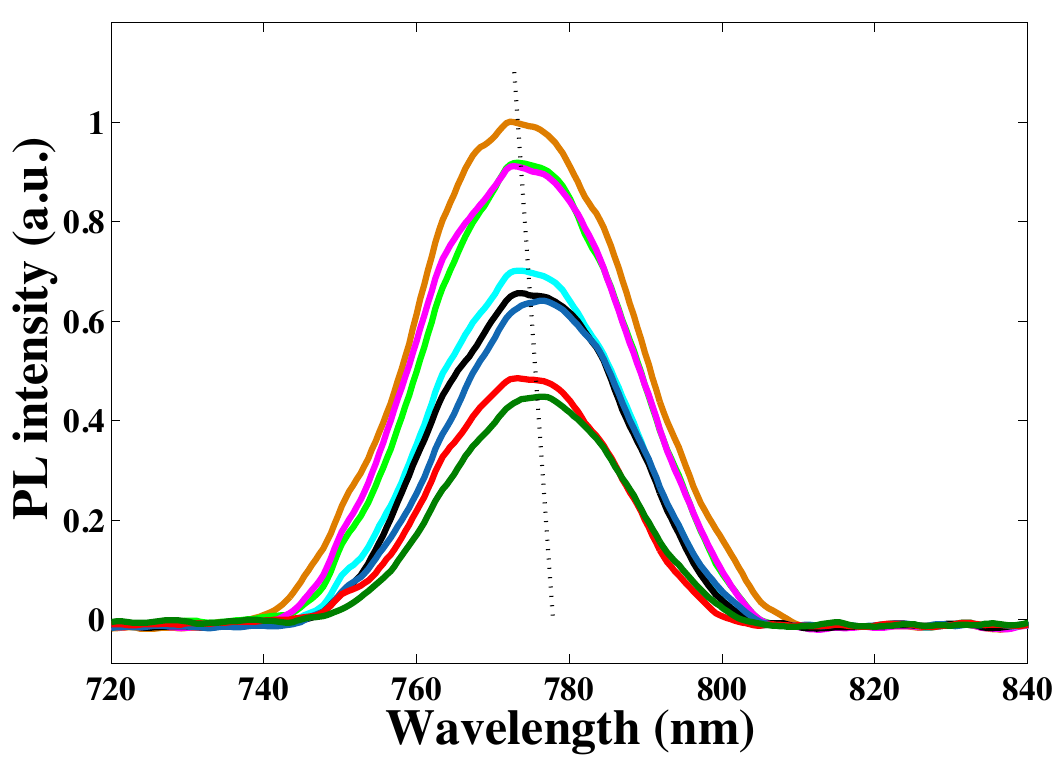}}
		\subfigure[]{\label{fig:9c} 
			\includegraphics[width=0.48\columnwidth]{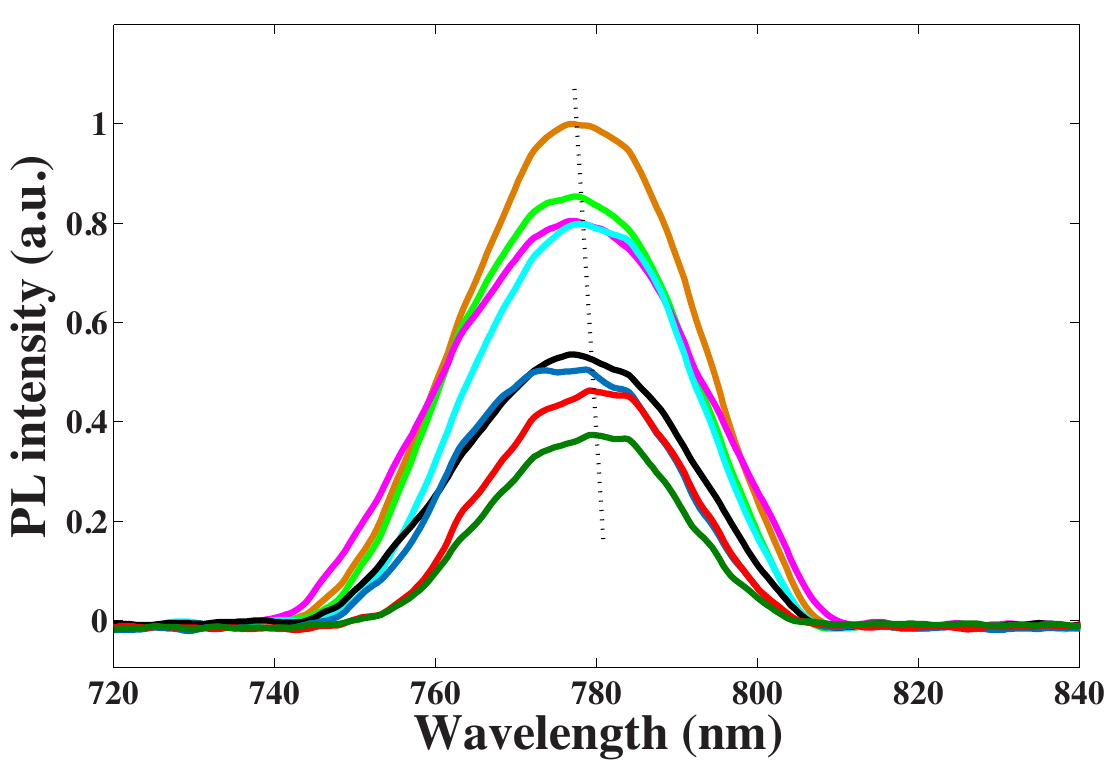}}
		\subfigure[]{\label{fig:10a} 
			\includegraphics[width=0.47\columnwidth]{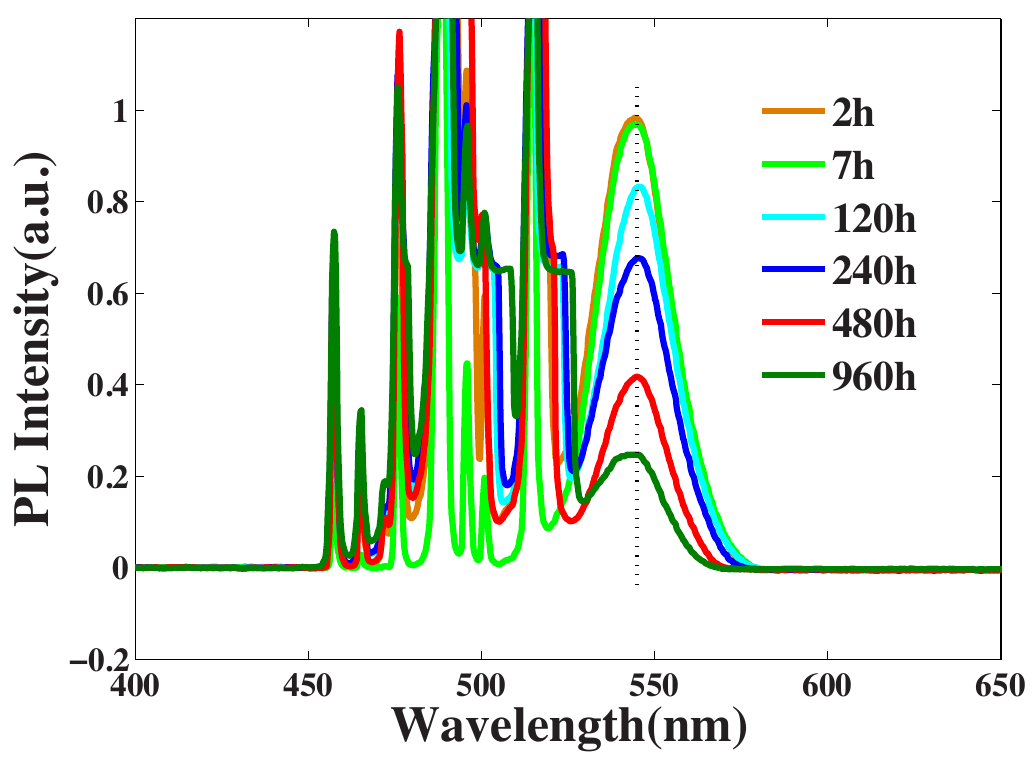}} 
		\subfigure[]{\label{fig:10b} 
			\includegraphics[width=0.47\columnwidth]{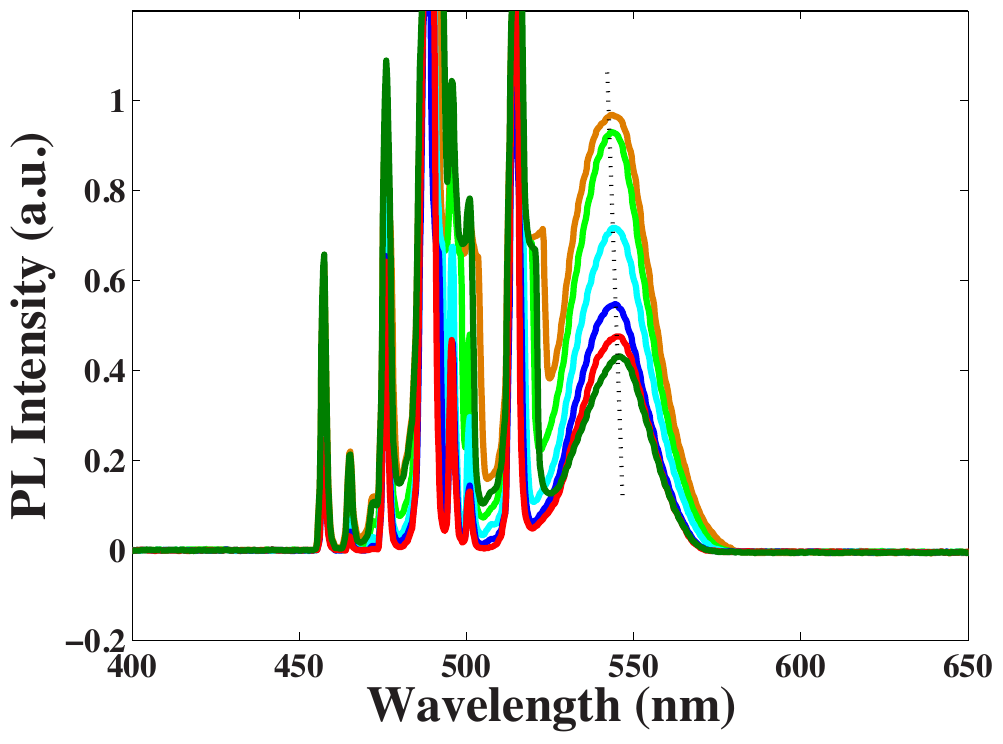}}
		\subfigure[]{\label{fig:10c} 
			\includegraphics[width=0.48\columnwidth]{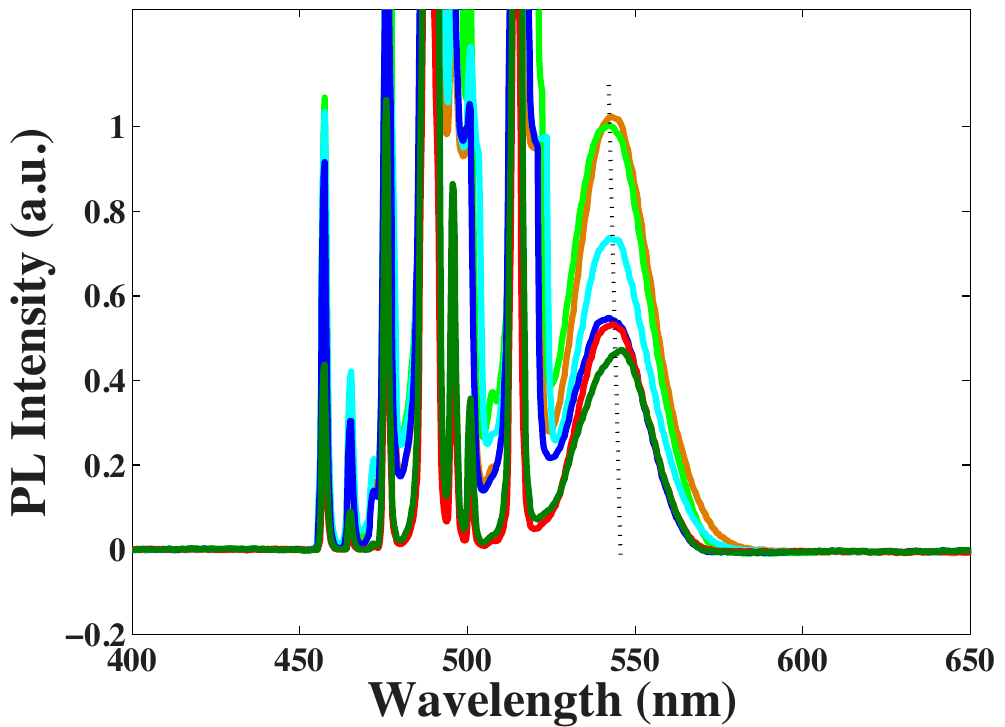}}
		\caption{PL spectra of (a) MAPbI$_3$ perovskite, (b) MAPbI$_3$ perovskite with PMMA, (c) MAPbI$_3$ perovskite with AZ5214 photoresist, (d) MAPbBr$_3$ perovskite, (e) MAPbBr$_3$ perovskite with PMMA, and (f) MAPbBr$_3$ perovskite with AZ5214 photoresist.} 
		\label{Fig:10}
	\end{center}
\end{figure}
\begin{figure}
	\begin{center}
		\subfigure[]{\label{fig:11a} 
			\includegraphics[width=0.48\columnwidth]{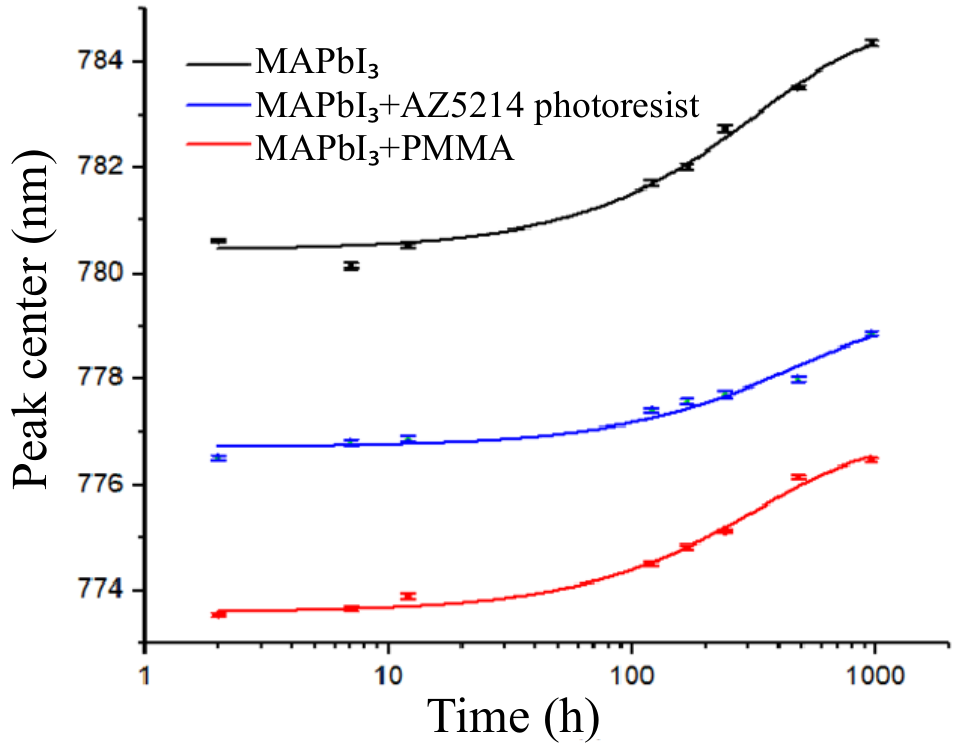}} 
		\subfigure[]{\label{fig:11b} 
			\includegraphics[width=0.48\columnwidth]{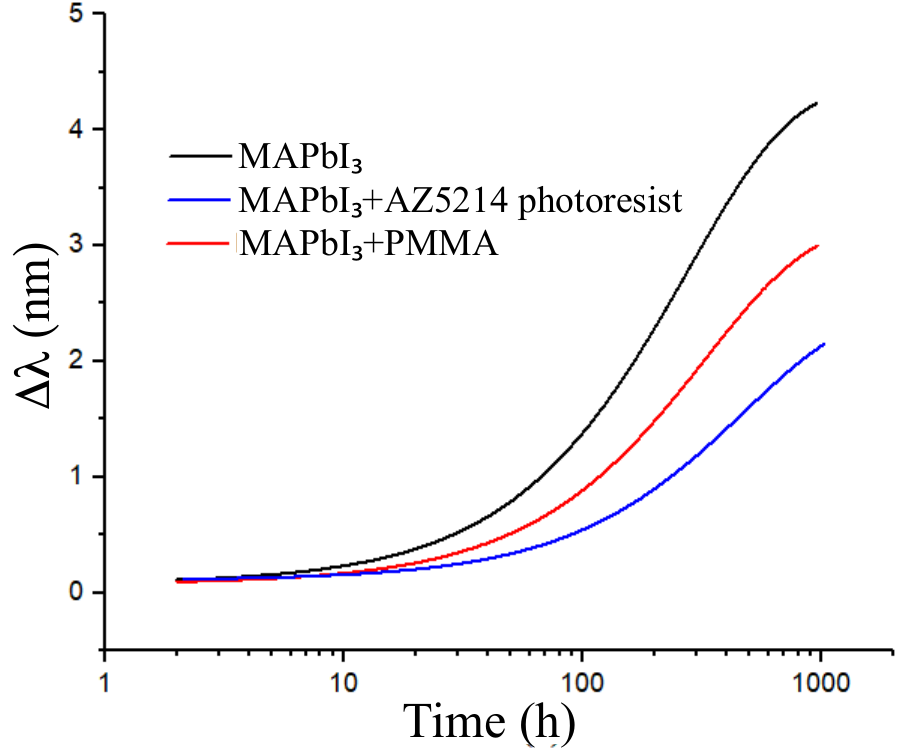}}
		\subfigure[]{\label{fig:12a} 
			\includegraphics[width=0.48\columnwidth]{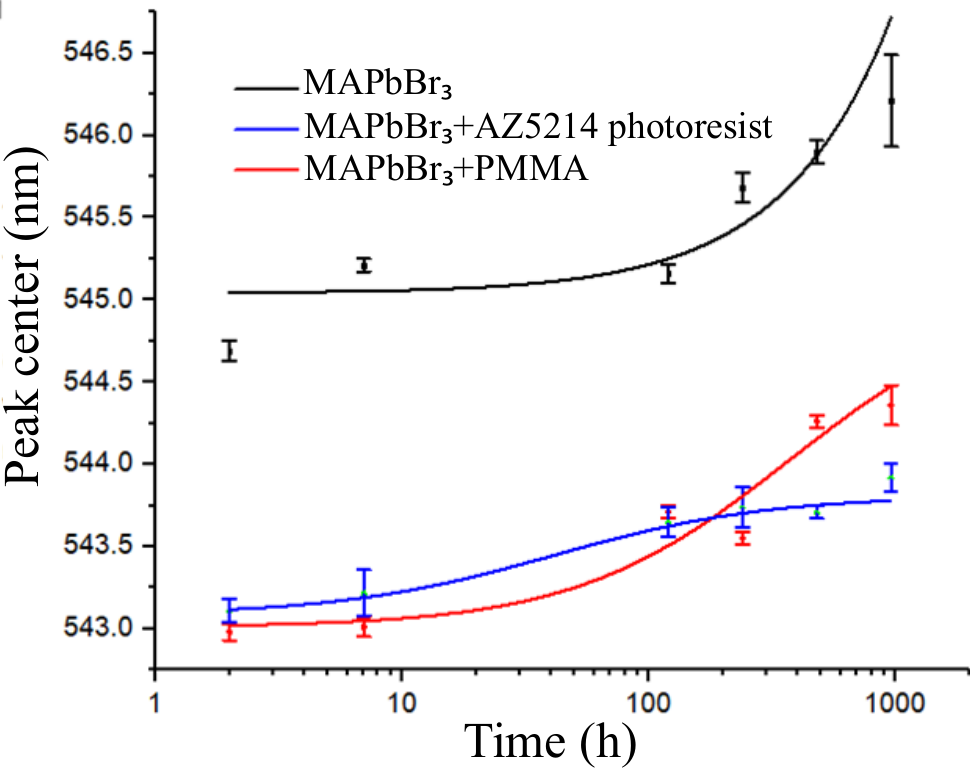}} 
		\subfigure[]{\label{fig:12b} 
			\includegraphics[width=0.48\columnwidth]{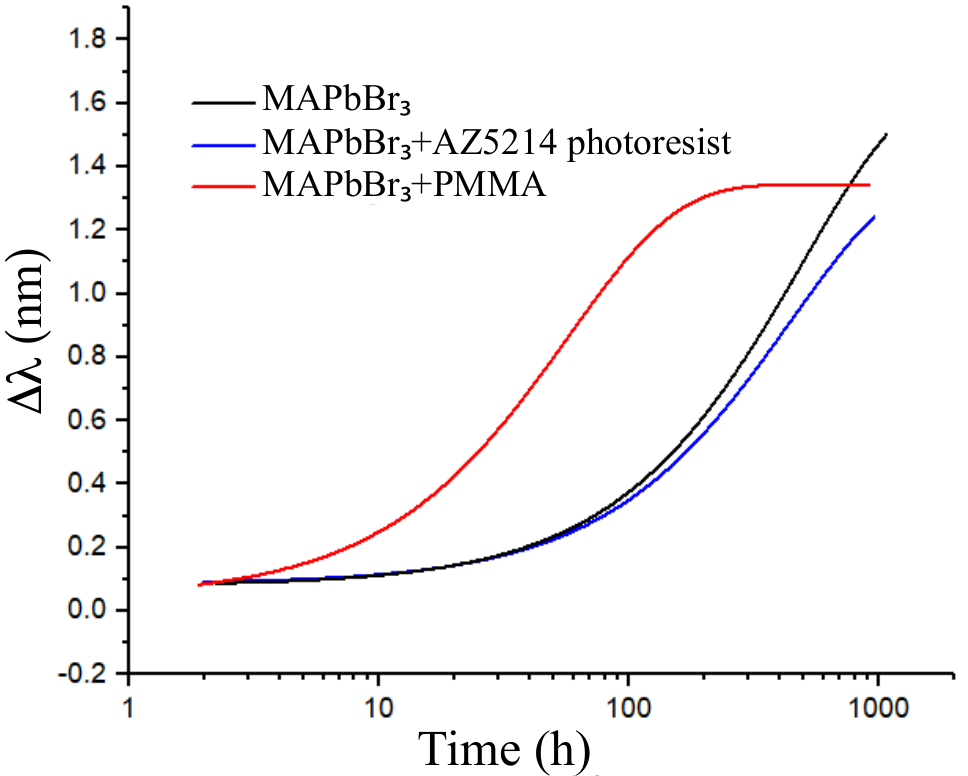}}
		\caption{Experimental results for peak center displacement and peak center displacement relative to the initial spectrum, i.e., $\Delta \lambda$. (a) peak center for MAPbI$_3$ perovskite, MAPbI$_3$ with AZ5214 photoresist, and MAPbI$_3$ with PMMA. (b) $\Delta \lambda$ for MAPbI$_3$ perovskite, MAPbI$_3$ with AZ5214 photoresist, and MAPbI$_3$ with PMMA. (c) peak center for MAPbBr$_3$ perovskite, MAPbBr$_3$ with AZ5214 photoresist, and MAPbBr$_3$ with PMMA. (d) $\Delta \lambda$ for MAPbBr$_3$ perovskite, MAPbBr$_3$ with AZ5214 photoresist, and MAPbBr$_3$ with PMMA.} 
		\label{Fig:12}
	\end{center}
\end{figure}

Figure 10 shows the experimental setup for photoluminescence (PL) spectroscopy.	Figures 11(a) to 11(f) depict the intensity of the PL spectra for MAPbI$_3$ and MAPbBr$_3$, respectively, showcasing optimized perovskite, perovskite with PMMA, and AZ5214 photoresist. In this process, we utilized laser emission wavelengths ranging from 457 to 514 nanometers. The emission of MAPbI$_3$ falls within a wavelength range of 750 to 850 nanometers, while that of MAPbBr$_3$ occurs between 560 and 575 nanometers. This discrepancy in wavelengths results in interference between the emissions. In the process of PL, electrons in the sample absorb energy from incident photons with energies higher than the bandgap of the electrons. As a result, electrons are excited to higher energy levels, and then they return to the ground state by emitting light again. Given that orthorhombic MAPbI$_3$ is a direct-band-gap crystal, the presence of multiple PL peaks signifies the involvement of various radiative recombination centers. These may include intrinsic defect states, extrinsic impurities, and residual tetragonal MAPbI$_3$ crystals. The deposition of PMMA and AZ5214 photoresist leads to a reduction in this red shift, which is one of the positive outcomes of this study. Figures 11(a) to 11(f) demonstrate that over time, the intensity of PL decreases for the optimized perovskite. However, after deposition with PMMA and AZ5214 photoresist, the decrease in PL intensity is less pronounced, indicating the stability of the perovskite layer.

In Figs. 12(a) to 12(d), the displacement of the peak center for the three states depicted in Figs. 11(a) to 11(f)  has been investigated for MAPbI$_3$ and MAPbBr$_3$ perovskites, respectively. The black curve corresponds to the perovskite with the maximum displacement of the peak center. One of the positive outcomes of employing PMMA and AZ5214 photoresist is the reduction of this peak displacement towards the right. Among these options, AZ5214 Photoresist proves to be superior to PMMA. This superiority stems from its reduced moisture penetration into the perovskite, thereby resulting in less structural degradation. In fact, the absence or reduction of a rightward shift serves as an indication of layer stability. With the coating of AZ5214 photoresist and PMMA on MAPbI$_3$, the peak center shifts by approximately 4 and 6 nanometers, respectively. Additionally, the rightward shift of the peak center will decrease. With the coating of these two materials on MAPbBr3 perovskite, the peak center shifts by approximately 2 nanometers (blue shift) and the shift amplitude is reduced. The positive results indicate that the use of AZ5214 photoresist and PMMA prevents shift, while the results related to AZ5214 photoresist are better than PMMA. The process of reducing the Full Width at Half Maximum (FWHM), the area under the curve, and the intensity with the coating of AZ5214 photoresist and PMMA on both perovskites decreases almost equally.

\newpage
\section{Discussion}\label{sec:}

The synthesized perovskite solutions, MAPbI$_3$ and MAPbBr$_3$, demonstrated efficient fabrication of thin films using the nanocrystal-pinning (NCP) process. Through SEM analysis, the morphology of the perovskite thin layer was elucidated, indicating successful encapsulation with PMMA and AZ5214 Photoresist without altering the inherent structure. This encapsulation process was further confirmed by XRD analysis, revealing enhanced stability of the perovskite layer, especially evident with AZ5214 Photoresist deposition. PL spectra analysis highlighted the sustained intensity of luminescence over time, emphasizing the role of PMMA and AZ5214 Photoresist in mitigating degradation. Importantly, the displacement of peak centers in the PL spectra indicated superior stability with AZ5214 Photoresist, attributable to its reduced moisture penetration and structural integrity. These findings underscore the novelty of utilizing AZ5214 Photoresist as a promising alternative for enhancing the durability and performance of perovskite-based optoelectronic devices.

Through our experimental investigations, we have demonstrated the effectiveness of matrix encapsulation in stabilizing perovskite thin films and prolonging their operational lifespan. Future research endeavors in this direction could explore novel encapsulating materials and deposition techniques to further enhance the performance and durability of perovskite-based devices for a wide range of applications in optoelectronics and beyond.

\section{Conclusions}\label{sec:} 
In conclusion, we have demonstrated the efficacy of AZ5214 photoresist deposition for stabilizing perovskite layers. Through a comparative analysis with PMMA deposition, AZ5214 photoresist proved superior in enhancing the stability of the perovskite layer, particularly at a thickness of 1.2 microns, surpassing the stabilization achieved with PMMA at a thickness of 127 nanometers. Moreover, the self-assembly properties of AZ5214 photoresist confer resistance against heat and moisture, further establishing its superiority over PMMA. These findings underscore the potential of AZ5214 photoresist as a promising candidate for enhancing the durability and performance of perovskite-based devices in various optoelectronic applications.

\bibliography{Taravati_Reference}

\begin{thebibliography}{37}%
\makeatletter
\providecommand \@ifxundefined [1]{%
 \@ifx{#1\undefined}
}%
\providecommand \@ifnum [1]{%
 \ifnum #1\expandafter \@firstoftwo
 \else \expandafter \@secondoftwo
 \fi
}%
\providecommand \@ifx [1]{%
 \ifx #1\expandafter \@firstoftwo
 \else \expandafter \@secondoftwo
 \fi
}%
\providecommand \natexlab [1]{#1}%
\providecommand \enquote  [1]{``#1''}%
\providecommand \bibnamefont  [1]{#1}%
\providecommand \bibfnamefont [1]{#1}%
\providecommand \citenamefont [1]{#1}%
\providecommand \href@noop [0]{\@secondoftwo}%
\providecommand \href [0]{\begingroup \@sanitize@url \@href}%
\providecommand \@href[1]{\@@startlink{#1}\@@href}%
\providecommand \@@href[1]{\endgroup#1\@@endlink}%
\providecommand \@sanitize@url [0]{\catcode `\\12\catcode `\$12\catcode
  `\&12\catcode `\#12\catcode `\^12\catcode `\_12\catcode `\%12\relax}%
\providecommand \@@startlink[1]{}%
\providecommand \@@endlink[0]{}%
\providecommand \url  [0]{\begingroup\@sanitize@url \@url }%
\providecommand \@url [1]{\endgroup\@href {#1}{\urlprefix }}%
\providecommand \urlprefix  [0]{URL }%
\providecommand \Eprint [0]{\href }%
\providecommand \doibase [0]{https://doi.org/}%
\providecommand \selectlanguage [0]{\@gobble}%
\providecommand \bibinfo  [0]{\@secondoftwo}%
\providecommand \bibfield  [0]{\@secondoftwo}%
\providecommand \translation [1]{[#1]}%
\providecommand \BibitemOpen [0]{}%
\providecommand \bibitemStop [0]{}%
\providecommand \bibitemNoStop [0]{.\EOS\space}%
\providecommand \EOS [0]{\spacefactor3000\relax}%
\providecommand \BibitemShut  [1]{\csname bibitem#1\endcsname}%
\let\auto@bib@innerbib\@empty
\bibitem [{\citenamefont {Wang}\ \emph {et~al.}(2024)\citenamefont {Wang},
  \citenamefont {Shu}, \citenamefont {Wei}, \citenamefont {Liang},
  \citenamefont {Ke}, \citenamefont {Shu},\ and\ \citenamefont
  {Catalan}}]{PhysRevLett.132.086902}%
  \BibitemOpen
  \bibfield  {author} {\bibinfo {author} {\bibfnamefont {Z.}~\bibnamefont
  {Wang}}, \bibinfo {author} {\bibfnamefont {S.}~\bibnamefont {Shu}}, \bibinfo
  {author} {\bibfnamefont {X.}~\bibnamefont {Wei}}, \bibinfo {author}
  {\bibfnamefont {R.}~\bibnamefont {Liang}}, \bibinfo {author} {\bibfnamefont
  {S.}~\bibnamefont {Ke}}, \bibinfo {author} {\bibfnamefont {L.}~\bibnamefont
  {Shu}},\ and\ \bibinfo {author} {\bibfnamefont {G.}~\bibnamefont {Catalan}},\
  }\bibfield  {title} {\bibinfo {title} {Flexophotovoltaic effect and
  above-band-gap photovoltage induced by strain gradients in halide
  perovskites},\ }\href {https://doi.org/10.1103/PhysRevLett.132.086902}
  {\bibfield  {journal} {\bibinfo  {journal} {Phys. Rev. Lett.}\ }\textbf
  {\bibinfo {volume} {132}},\ \bibinfo {pages} {086902} (\bibinfo {year}
  {2024})}\BibitemShut {NoStop}%
\bibitem [{\citenamefont {Kangsabanik}\ \emph {et~al.}(2020)\citenamefont
  {Kangsabanik}, \citenamefont {Ghorui}, \citenamefont {Aslam},\ and\
  \citenamefont {Alam}}]{PhysRevApplied.13.014005}%
  \BibitemOpen
  \bibfield  {author} {\bibinfo {author} {\bibfnamefont {J.}~\bibnamefont
  {Kangsabanik}}, \bibinfo {author} {\bibfnamefont {S.}~\bibnamefont {Ghorui}},
  \bibinfo {author} {\bibfnamefont {M.}~\bibnamefont {Aslam}},\ and\ \bibinfo
  {author} {\bibfnamefont {A.}~\bibnamefont {Alam}},\ }\bibfield  {title}
  {\bibinfo {title} {Optoelectronic properties and defect physics of lead-free
  photovoltaic absorbers
  ${\mathrm{cs}}_{2}{\mathrm{au}}^{\mathrm{i}}{\mathrm{au}}^{\mathrm{iii}}{X}_{6}$
  ($x=\mathrm{I},\phantom{\rule{0.2em}{0ex}}\mathrm{Br}$)},\ }\href
  {https://doi.org/10.1103/PhysRevApplied.13.014005} {\bibfield  {journal}
  {\bibinfo  {journal} {Phys. Rev. Appl.}\ }\textbf {\bibinfo {volume} {13}},\
  \bibinfo {pages} {014005} (\bibinfo {year} {2020})}\BibitemShut {NoStop}%
\bibitem [{\citenamefont {Ferreira}\ \emph {et~al.}(2018)\citenamefont
  {Ferreira}, \citenamefont {L\'etoublon}, \citenamefont {Paofai},
  \citenamefont {Raymond}, \citenamefont {Ecolivet}, \citenamefont {Ruffl\'e},
  \citenamefont {Cordier}, \citenamefont {Katan}, \citenamefont {Saidaminov},
  \citenamefont {Zhumekenov}, \citenamefont {Bakr}, \citenamefont {Even},\ and\
  \citenamefont {Bourges}}]{PhysRevLett.121.085502}%
  \BibitemOpen
  \bibfield  {author} {\bibinfo {author} {\bibfnamefont {A.~C.}\ \bibnamefont
  {Ferreira}}, \bibinfo {author} {\bibfnamefont {A.}~\bibnamefont
  {L\'etoublon}}, \bibinfo {author} {\bibfnamefont {S.}~\bibnamefont {Paofai}},
  \bibinfo {author} {\bibfnamefont {S.}~\bibnamefont {Raymond}}, \bibinfo
  {author} {\bibfnamefont {C.}~\bibnamefont {Ecolivet}}, \bibinfo {author}
  {\bibfnamefont {B.}~\bibnamefont {Ruffl\'e}}, \bibinfo {author}
  {\bibfnamefont {S.}~\bibnamefont {Cordier}}, \bibinfo {author} {\bibfnamefont
  {C.}~\bibnamefont {Katan}}, \bibinfo {author} {\bibfnamefont {M.~I.}\
  \bibnamefont {Saidaminov}}, \bibinfo {author} {\bibfnamefont {A.~A.}\
  \bibnamefont {Zhumekenov}}, \bibinfo {author} {\bibfnamefont {O.~M.}\
  \bibnamefont {Bakr}}, \bibinfo {author} {\bibfnamefont {J.}~\bibnamefont
  {Even}},\ and\ \bibinfo {author} {\bibfnamefont {P.}~\bibnamefont
  {Bourges}},\ }\bibfield  {title} {\bibinfo {title} {Elastic softness of
  hybrid lead halide perovskites},\ }\href
  {https://doi.org/10.1103/PhysRevLett.121.085502} {\bibfield  {journal}
  {\bibinfo  {journal} {Phys. Rev. Lett.}\ }\textbf {\bibinfo {volume} {121}},\
  \bibinfo {pages} {085502} (\bibinfo {year} {2018})}\BibitemShut {NoStop}%
\bibitem [{\citenamefont {de~Quilettes}\ \emph {et~al.}(2015)\citenamefont
  {de~Quilettes}, \citenamefont {Vorpahl}, \citenamefont {Stranks},
  \citenamefont {Nagaoka}, \citenamefont {Eperon}, \citenamefont {Ziffer},
  \citenamefont {Snaith},\ and\ \citenamefont {Ginger}}]{de2015impact}%
  \BibitemOpen
  \bibfield  {author} {\bibinfo {author} {\bibfnamefont {D.~W.}\ \bibnamefont
  {de~Quilettes}}, \bibinfo {author} {\bibfnamefont {S.~M.}\ \bibnamefont
  {Vorpahl}}, \bibinfo {author} {\bibfnamefont {S.~D.}\ \bibnamefont
  {Stranks}}, \bibinfo {author} {\bibfnamefont {H.}~\bibnamefont {Nagaoka}},
  \bibinfo {author} {\bibfnamefont {G.~E.}\ \bibnamefont {Eperon}}, \bibinfo
  {author} {\bibfnamefont {M.~E.}\ \bibnamefont {Ziffer}}, \bibinfo {author}
  {\bibfnamefont {H.~J.}\ \bibnamefont {Snaith}},\ and\ \bibinfo {author}
  {\bibfnamefont {D.~S.}\ \bibnamefont {Ginger}},\ }\bibfield  {title}
  {\bibinfo {title} {Impact of microstructure on local carrier lifetime in
  perovskite solar cells},\ }\href@noop {} {\bibfield  {journal} {\bibinfo
  {journal} {Science}\ }\textbf {\bibinfo {volume} {348}},\ \bibinfo {pages}
  {683} (\bibinfo {year} {2015})}\BibitemShut {NoStop}%
\bibitem [{\citenamefont {Jeon}\ \emph {et~al.}(2018)\citenamefont {Jeon},
  \citenamefont {Na}, \citenamefont {Jung}, \citenamefont {Yang}, \citenamefont
  {Lee}, \citenamefont {Kim}, \citenamefont {Shin}, \citenamefont {Il~Seok},
  \citenamefont {Lee},\ and\ \citenamefont {Seo}}]{jeon2018fluorene}%
  \BibitemOpen
  \bibfield  {author} {\bibinfo {author} {\bibfnamefont {N.~J.}\ \bibnamefont
  {Jeon}}, \bibinfo {author} {\bibfnamefont {H.}~\bibnamefont {Na}}, \bibinfo
  {author} {\bibfnamefont {E.~H.}\ \bibnamefont {Jung}}, \bibinfo {author}
  {\bibfnamefont {T.-Y.}\ \bibnamefont {Yang}}, \bibinfo {author}
  {\bibfnamefont {Y.~G.}\ \bibnamefont {Lee}}, \bibinfo {author} {\bibfnamefont
  {G.}~\bibnamefont {Kim}}, \bibinfo {author} {\bibfnamefont {H.-W.}\
  \bibnamefont {Shin}}, \bibinfo {author} {\bibfnamefont {S.}~\bibnamefont
  {Il~Seok}}, \bibinfo {author} {\bibfnamefont {J.}~\bibnamefont {Lee}},\ and\
  \bibinfo {author} {\bibfnamefont {J.}~\bibnamefont {Seo}},\ }\bibfield
  {title} {\bibinfo {title} {A fluorene-terminated hole-transporting material
  for highly efficient and stable perovskite solar cells},\ }\href@noop {}
  {\bibfield  {journal} {\bibinfo  {journal} {Nature Energy}\ }\textbf
  {\bibinfo {volume} {3}},\ \bibinfo {pages} {682} (\bibinfo {year}
  {2018})}\BibitemShut {NoStop}%
\bibitem [{\citenamefont {Brenner}\ \emph {et~al.}(2016)\citenamefont
  {Brenner}, \citenamefont {Egger}, \citenamefont {Kronik}, \citenamefont
  {Hodes},\ and\ \citenamefont {Cahen}}]{brenner2016hybrid}%
  \BibitemOpen
  \bibfield  {author} {\bibinfo {author} {\bibfnamefont {T.~M.}\ \bibnamefont
  {Brenner}}, \bibinfo {author} {\bibfnamefont {D.~A.}\ \bibnamefont {Egger}},
  \bibinfo {author} {\bibfnamefont {L.}~\bibnamefont {Kronik}}, \bibinfo
  {author} {\bibfnamefont {G.}~\bibnamefont {Hodes}},\ and\ \bibinfo {author}
  {\bibfnamefont {D.}~\bibnamefont {Cahen}},\ }\bibfield  {title} {\bibinfo
  {title} {Hybrid organic—inorganic perovskites: low-cost semiconductors with
  intriguing charge-transport properties},\ }\href@noop {} {\bibfield
  {journal} {\bibinfo  {journal} {Nature Reviews Materials}\ }\textbf {\bibinfo
  {volume} {1}},\ \bibinfo {pages} {1} (\bibinfo {year} {2016})}\BibitemShut
  {NoStop}%
\bibitem [{\citenamefont {Qin}\ \emph {et~al.}(2014)\citenamefont {Qin},
  \citenamefont {Tanaka}, \citenamefont {Ito}, \citenamefont {Tetreault},
  \citenamefont {Manabe}, \citenamefont {Nishino}, \citenamefont
  {Nazeeruddin},\ and\ \citenamefont {Gr{\"a}tzel}}]{qin2014inorganic}%
  \BibitemOpen
  \bibfield  {author} {\bibinfo {author} {\bibfnamefont {P.}~\bibnamefont
  {Qin}}, \bibinfo {author} {\bibfnamefont {S.}~\bibnamefont {Tanaka}},
  \bibinfo {author} {\bibfnamefont {S.}~\bibnamefont {Ito}}, \bibinfo {author}
  {\bibfnamefont {N.}~\bibnamefont {Tetreault}}, \bibinfo {author}
  {\bibfnamefont {K.}~\bibnamefont {Manabe}}, \bibinfo {author} {\bibfnamefont
  {H.}~\bibnamefont {Nishino}}, \bibinfo {author} {\bibfnamefont {M.~K.}\
  \bibnamefont {Nazeeruddin}},\ and\ \bibinfo {author} {\bibfnamefont
  {M.}~\bibnamefont {Gr{\"a}tzel}},\ }\bibfield  {title} {\bibinfo {title}
  {Inorganic hole conductor-based lead halide perovskite solar cells with
  12.4\% conversion efficiency},\ }\href@noop {} {\bibfield  {journal}
  {\bibinfo  {journal} {Nature communications}\ }\textbf {\bibinfo {volume}
  {5}},\ \bibinfo {pages} {3834} (\bibinfo {year} {2014})}\BibitemShut
  {NoStop}%
\bibitem [{\citenamefont {Basera}\ \emph {et~al.}(2020)\citenamefont {Basera},
  \citenamefont {Kumar}, \citenamefont {Saini},\ and\ \citenamefont
  {Bhattacharya}}]{PhysRevB.101.054108}%
  \BibitemOpen
  \bibfield  {author} {\bibinfo {author} {\bibfnamefont {P.}~\bibnamefont
  {Basera}}, \bibinfo {author} {\bibfnamefont {M.}~\bibnamefont {Kumar}},
  \bibinfo {author} {\bibfnamefont {S.}~\bibnamefont {Saini}},\ and\ \bibinfo
  {author} {\bibfnamefont {S.}~\bibnamefont {Bhattacharya}},\ }\bibfield
  {title} {\bibinfo {title} {Reducing lead toxicity in the methylammonium lead
  halide ${\mathrm{mapbi}}_{3}$: Why sn substitution should be preferred to pb
  vacancy for optimum solar cell efficiency},\ }\href
  {https://doi.org/10.1103/PhysRevB.101.054108} {\bibfield  {journal} {\bibinfo
   {journal} {Phys. Rev. B}\ }\textbf {\bibinfo {volume} {101}},\ \bibinfo
  {pages} {054108} (\bibinfo {year} {2020})}\BibitemShut {NoStop}%
\bibitem [{\citenamefont {Whitcher}\ \emph {et~al.}(2018)\citenamefont
  {Whitcher}, \citenamefont {Zhu}, \citenamefont {Chi}, \citenamefont {Hu},
  \citenamefont {Zhao}, \citenamefont {Asmara}, \citenamefont {Yu},
  \citenamefont {Breese}, \citenamefont {Castro~Neto}, \citenamefont {Lam},
  \citenamefont {Wee}, \citenamefont {Chia},\ and\ \citenamefont
  {Rusydi}}]{PhysRevX.8.021034}%
  \BibitemOpen
  \bibfield  {author} {\bibinfo {author} {\bibfnamefont {T.~J.}\ \bibnamefont
  {Whitcher}}, \bibinfo {author} {\bibfnamefont {J.-X.}\ \bibnamefont {Zhu}},
  \bibinfo {author} {\bibfnamefont {X.}~\bibnamefont {Chi}}, \bibinfo {author}
  {\bibfnamefont {H.}~\bibnamefont {Hu}}, \bibinfo {author} {\bibfnamefont
  {D.}~\bibnamefont {Zhao}}, \bibinfo {author} {\bibfnamefont {T.~C.}\
  \bibnamefont {Asmara}}, \bibinfo {author} {\bibfnamefont {X.}~\bibnamefont
  {Yu}}, \bibinfo {author} {\bibfnamefont {M.~B.~H.}\ \bibnamefont {Breese}},
  \bibinfo {author} {\bibfnamefont {A.~H.}\ \bibnamefont {Castro~Neto}},
  \bibinfo {author} {\bibfnamefont {Y.~M.}\ \bibnamefont {Lam}}, \bibinfo
  {author} {\bibfnamefont {A.~T.~S.}\ \bibnamefont {Wee}}, \bibinfo {author}
  {\bibfnamefont {E.~E.~M.}\ \bibnamefont {Chia}},\ and\ \bibinfo {author}
  {\bibfnamefont {A.}~\bibnamefont {Rusydi}},\ }\bibfield  {title} {\bibinfo
  {title} {Importance of electronic correlations and unusual excitonic effects
  in formamidinium lead halide perovskites},\ }\href
  {https://doi.org/10.1103/PhysRevX.8.021034} {\bibfield  {journal} {\bibinfo
  {journal} {Phys. Rev. X}\ }\textbf {\bibinfo {volume} {8}},\ \bibinfo {pages}
  {021034} (\bibinfo {year} {2018})}\BibitemShut {NoStop}%
\bibitem [{\citenamefont {Akkerman}\ \emph {et~al.}(2018)\citenamefont
  {Akkerman}, \citenamefont {Rain{\`o}}, \citenamefont {Kovalenko},\ and\
  \citenamefont {Manna}}]{akkerman2018genesis}%
  \BibitemOpen
  \bibfield  {author} {\bibinfo {author} {\bibfnamefont {Q.~A.}\ \bibnamefont
  {Akkerman}}, \bibinfo {author} {\bibfnamefont {G.}~\bibnamefont {Rain{\`o}}},
  \bibinfo {author} {\bibfnamefont {M.~V.}\ \bibnamefont {Kovalenko}},\ and\
  \bibinfo {author} {\bibfnamefont {L.}~\bibnamefont {Manna}},\ }\bibfield
  {title} {\bibinfo {title} {Genesis, challenges and opportunities for
  colloidal lead halide perovskite nanocrystals},\ }\href@noop {} {\bibfield
  {journal} {\bibinfo  {journal} {Nature materials}\ }\textbf {\bibinfo
  {volume} {17}},\ \bibinfo {pages} {394} (\bibinfo {year} {2018})}\BibitemShut
  {NoStop}%
\bibitem [{\citenamefont {Xue}\ \emph {et~al.}(2022)\citenamefont {Xue},
  \citenamefont {Brocks},\ and\ \citenamefont
  {Tao}}]{PhysRevMaterials.6.055402}%
  \BibitemOpen
  \bibfield  {author} {\bibinfo {author} {\bibfnamefont {H.}~\bibnamefont
  {Xue}}, \bibinfo {author} {\bibfnamefont {G.}~\bibnamefont {Brocks}},\ and\
  \bibinfo {author} {\bibfnamefont {S.}~\bibnamefont {Tao}},\ }\bibfield
  {title} {\bibinfo {title} {Intrinsic defects in primary halide perovskites: A
  first-principles study of the thermodynamic trends},\ }\href
  {https://doi.org/10.1103/PhysRevMaterials.6.055402} {\bibfield  {journal}
  {\bibinfo  {journal} {Phys. Rev. Mater.}\ }\textbf {\bibinfo {volume} {6}},\
  \bibinfo {pages} {055402} (\bibinfo {year} {2022})}\BibitemShut {NoStop}%
\bibitem [{\citenamefont {Stranks}\ and\ \citenamefont
  {Snaith}(2015)}]{stranks2015metal}%
  \BibitemOpen
  \bibfield  {author} {\bibinfo {author} {\bibfnamefont {S.~D.}\ \bibnamefont
  {Stranks}}\ and\ \bibinfo {author} {\bibfnamefont {H.~J.}\ \bibnamefont
  {Snaith}},\ }\bibfield  {title} {\bibinfo {title} {Metal-halide perovskites
  for photovoltaic and light-emitting devices},\ }\href@noop {} {\bibfield
  {journal} {\bibinfo  {journal} {Nature nanotechnology}\ }\textbf {\bibinfo
  {volume} {10}},\ \bibinfo {pages} {391} (\bibinfo {year} {2015})}\BibitemShut
  {NoStop}%
\bibitem [{\citenamefont {Taravati}\ and\ \citenamefont
  {Eleftheriades}(2019)}]{PhysRevApplied.12.024026}%
  \BibitemOpen
  \bibfield  {author} {\bibinfo {author} {\bibfnamefont {S.}~\bibnamefont
  {Taravati}}\ and\ \bibinfo {author} {\bibfnamefont {G.~V.}\ \bibnamefont
  {Eleftheriades}},\ }\bibfield  {title} {\bibinfo {title} {Generalized
  space-time-periodic diffraction gratings: Theory and applications},\ }\href
  {https://doi.org/10.1103/PhysRevApplied.12.024026} {\bibfield  {journal}
  {\bibinfo  {journal} {Phys. Rev. Appl.}\ }\textbf {\bibinfo {volume} {12}},\
  \bibinfo {pages} {024026} (\bibinfo {year} {2019})}\BibitemShut {NoStop}%
\bibitem [{\citenamefont {Taravati}\ and\ \citenamefont
  {Eleftheriades}(2020)}]{PhysRevApplied.14.014027}%
  \BibitemOpen
  \bibfield  {author} {\bibinfo {author} {\bibfnamefont {S.}~\bibnamefont
  {Taravati}}\ and\ \bibinfo {author} {\bibfnamefont {G.~V.}\ \bibnamefont
  {Eleftheriades}},\ }\bibfield  {title} {\bibinfo {title} {Full-duplex
  nonreciprocal beam steering by time-modulated phase-gradient metasurfaces},\
  }\href {https://doi.org/10.1103/PhysRevApplied.14.014027} {\bibfield
  {journal} {\bibinfo  {journal} {Phys. Rev. Appl.}\ }\textbf {\bibinfo
  {volume} {14}},\ \bibinfo {pages} {014027} (\bibinfo {year}
  {2020})}\BibitemShut {NoStop}%
\bibitem [{\citenamefont {Taravati}\ and\ \citenamefont
  {Eleftheriades}(2021)}]{Taravati_NC_2021}%
  \BibitemOpen
  \bibfield  {author} {\bibinfo {author} {\bibfnamefont {S.}~\bibnamefont
  {Taravati}}\ and\ \bibinfo {author} {\bibfnamefont {G.~V.}\ \bibnamefont
  {Eleftheriades}},\ }\bibfield  {title} {\bibinfo {title} {Full-duplex
  reflective beamsteering metasurface featuring magnetless nonreciprocal
  amplification},\ }\href@noop {} {\bibfield  {journal} {\bibinfo  {journal}
  {{Nat. Commun.}}\ }\textbf {\bibinfo {volume} {14}},\ \bibinfo {pages} {4414}
  (\bibinfo {year} {2021})}\BibitemShut {NoStop}%
\bibitem [{\citenamefont {Taravati}\ and\ \citenamefont
  {Eleftheriades}(2022)}]{Taravati_ACSP_2022}%
  \BibitemOpen
  \bibfield  {author} {\bibinfo {author} {\bibfnamefont {S.}~\bibnamefont
  {Taravati}}\ and\ \bibinfo {author} {\bibfnamefont {G.~V.}\ \bibnamefont
  {Eleftheriades}},\ }\bibfield  {title} {\bibinfo {title} {Microwave
  space-time-modulated metasurfaces},\ }\href@noop {} {\bibfield  {journal}
  {\bibinfo  {journal} {{ACS Photonics}}\ }\textbf {\bibinfo {volume} {9}},\
  \bibinfo {pages} {305} (\bibinfo {year} {2022})}\BibitemShut {NoStop}%
\bibitem [{\citenamefont {As'ham}\ \emph {et~al.}(2022)\citenamefont {As'ham},
  \citenamefont {Al-Ani}, \citenamefont {Lei}, \citenamefont {Hattori},
  \citenamefont {Huang},\ and\ \citenamefont
  {Miroshnichenko}}]{PhysRevApplied.18.014079}%
  \BibitemOpen
  \bibfield  {author} {\bibinfo {author} {\bibfnamefont {K.}~\bibnamefont
  {As'ham}}, \bibinfo {author} {\bibfnamefont {I.}~\bibnamefont {Al-Ani}},
  \bibinfo {author} {\bibfnamefont {W.}~\bibnamefont {Lei}}, \bibinfo {author}
  {\bibfnamefont {H.~T.}\ \bibnamefont {Hattori}}, \bibinfo {author}
  {\bibfnamefont {L.}~\bibnamefont {Huang}},\ and\ \bibinfo {author}
  {\bibfnamefont {A.}~\bibnamefont {Miroshnichenko}},\ }\bibfield  {title}
  {\bibinfo {title} {Mie exciton-polariton in a perovskite metasurface},\
  }\href {https://doi.org/10.1103/PhysRevApplied.18.014079} {\bibfield
  {journal} {\bibinfo  {journal} {Phys. Rev. Appl.}\ }\textbf {\bibinfo
  {volume} {18}},\ \bibinfo {pages} {014079} (\bibinfo {year}
  {2022})}\BibitemShut {NoStop}%
\bibitem [{\citenamefont {Ye}\ \emph {et~al.}(2016)\citenamefont {Ye},
  \citenamefont {Chen}, \citenamefont {Xie}, \citenamefont {Tang},
  \citenamefont {Yin}, \citenamefont {He}, \citenamefont {Bi}, \citenamefont
  {Wang}, \citenamefont {Yang},\ and\ \citenamefont {Han}}]{ye2016soft}%
  \BibitemOpen
  \bibfield  {author} {\bibinfo {author} {\bibfnamefont {F.}~\bibnamefont
  {Ye}}, \bibinfo {author} {\bibfnamefont {H.}~\bibnamefont {Chen}}, \bibinfo
  {author} {\bibfnamefont {F.}~\bibnamefont {Xie}}, \bibinfo {author}
  {\bibfnamefont {W.}~\bibnamefont {Tang}}, \bibinfo {author} {\bibfnamefont
  {M.}~\bibnamefont {Yin}}, \bibinfo {author} {\bibfnamefont {J.}~\bibnamefont
  {He}}, \bibinfo {author} {\bibfnamefont {E.}~\bibnamefont {Bi}}, \bibinfo
  {author} {\bibfnamefont {Y.}~\bibnamefont {Wang}}, \bibinfo {author}
  {\bibfnamefont {X.}~\bibnamefont {Yang}},\ and\ \bibinfo {author}
  {\bibfnamefont {L.}~\bibnamefont {Han}},\ }\bibfield  {title} {\bibinfo
  {title} {Soft-cover deposition of scaling-up uniform perovskite thin films
  for high cost-performance solar cells},\ }\href@noop {} {\bibfield  {journal}
  {\bibinfo  {journal} {Energy \& Environmental Science}\ }\textbf {\bibinfo
  {volume} {9}},\ \bibinfo {pages} {2295} (\bibinfo {year} {2016})}\BibitemShut
  {NoStop}%
\bibitem [{\citenamefont {Li}\ \emph {et~al.}(2018{\natexlab{a}})\citenamefont
  {Li}, \citenamefont {Zhao}, \citenamefont {Yang}, \citenamefont {Du},\ and\
  \citenamefont {Zhang}}]{PhysRevApplied.10.041001}%
  \BibitemOpen
  \bibfield  {author} {\bibinfo {author} {\bibfnamefont {T.}~\bibnamefont
  {Li}}, \bibinfo {author} {\bibfnamefont {X.}~\bibnamefont {Zhao}}, \bibinfo
  {author} {\bibfnamefont {D.}~\bibnamefont {Yang}}, \bibinfo {author}
  {\bibfnamefont {M.-H.}\ \bibnamefont {Du}},\ and\ \bibinfo {author}
  {\bibfnamefont {L.}~\bibnamefont {Zhang}},\ }\bibfield  {title} {\bibinfo
  {title} {Intrinsic defect properties in halide double perovskites for
  optoelectronic applications},\ }\href
  {https://doi.org/10.1103/PhysRevApplied.10.041001} {\bibfield  {journal}
  {\bibinfo  {journal} {Phys. Rev. Appl.}\ }\textbf {\bibinfo {volume} {10}},\
  \bibinfo {pages} {041001} (\bibinfo {year} {2018}{\natexlab{a}})}\BibitemShut
  {NoStop}%
\bibitem [{\citenamefont {Green}\ \emph {et~al.}(2014)\citenamefont {Green},
  \citenamefont {Ho-Baillie},\ and\ \citenamefont
  {Snaith}}]{green2014emergence}%
  \BibitemOpen
  \bibfield  {author} {\bibinfo {author} {\bibfnamefont {M.~A.}\ \bibnamefont
  {Green}}, \bibinfo {author} {\bibfnamefont {A.}~\bibnamefont {Ho-Baillie}},\
  and\ \bibinfo {author} {\bibfnamefont {H.~J.}\ \bibnamefont {Snaith}},\
  }\bibfield  {title} {\bibinfo {title} {The emergence of perovskite solar
  cells},\ }\href@noop {} {\bibfield  {journal} {\bibinfo  {journal} {Nature
  photonics}\ }\textbf {\bibinfo {volume} {8}},\ \bibinfo {pages} {506}
  (\bibinfo {year} {2014})}\BibitemShut {NoStop}%
\bibitem [{\citenamefont {Akhanuly}\ \emph {et~al.}(2023)\citenamefont
  {Akhanuly}, \citenamefont {Dossyaev}, \citenamefont {Shalenov}, \citenamefont
  {Valagiannopoulos}, \citenamefont {Dzhumagulova}, \citenamefont {Ng},\ and\
  \citenamefont {Jumabekov}}]{PhysRevApplied.19.054039}%
  \BibitemOpen
  \bibfield  {author} {\bibinfo {author} {\bibfnamefont {A.}~\bibnamefont
  {Akhanuly}}, \bibinfo {author} {\bibfnamefont {I.~T.}\ \bibnamefont
  {Dossyaev}}, \bibinfo {author} {\bibfnamefont {E.~O.}\ \bibnamefont
  {Shalenov}}, \bibinfo {author} {\bibfnamefont {C.}~\bibnamefont
  {Valagiannopoulos}}, \bibinfo {author} {\bibfnamefont {K.~N.}\ \bibnamefont
  {Dzhumagulova}}, \bibinfo {author} {\bibfnamefont {A.}~\bibnamefont {Ng}},\
  and\ \bibinfo {author} {\bibfnamefont {A.~N.}\ \bibnamefont {Jumabekov}},\
  }\bibfield  {title} {\bibinfo {title} {Modeling and comparative performance
  analysis of perovskite solar cells with planar or nanorod
  ${\mathrm{sn}\mathrm{o}}_{2}$ electron-transport layers},\ }\href
  {https://doi.org/10.1103/PhysRevApplied.19.054039} {\bibfield  {journal}
  {\bibinfo  {journal} {Phys. Rev. Appl.}\ }\textbf {\bibinfo {volume} {19}},\
  \bibinfo {pages} {054039} (\bibinfo {year} {2023})}\BibitemShut {NoStop}%
\bibitem [{\citenamefont {Jiang}\ \emph {et~al.}(2018)\citenamefont {Jiang},
  \citenamefont {Liu}, \citenamefont {Su}, \citenamefont {Yu}, \citenamefont
  {Xu}, \citenamefont {Wei}, \citenamefont {Zhou},\ and\ \citenamefont
  {Wang}}]{jiang2018continuous}%
  \BibitemOpen
  \bibfield  {author} {\bibinfo {author} {\bibfnamefont {L.}~\bibnamefont
  {Jiang}}, \bibinfo {author} {\bibfnamefont {R.}~\bibnamefont {Liu}}, \bibinfo
  {author} {\bibfnamefont {R.}~\bibnamefont {Su}}, \bibinfo {author}
  {\bibfnamefont {Y.}~\bibnamefont {Yu}}, \bibinfo {author} {\bibfnamefont
  {H.}~\bibnamefont {Xu}}, \bibinfo {author} {\bibfnamefont {Y.}~\bibnamefont
  {Wei}}, \bibinfo {author} {\bibfnamefont {Z.-K.}\ \bibnamefont {Zhou}},\ and\
  \bibinfo {author} {\bibfnamefont {X.}~\bibnamefont {Wang}},\ }\bibfield
  {title} {\bibinfo {title} {Continuous wave pumped single-mode nanolasers in
  inorganic perovskites with robust stability and high quantum yield},\
  }\href@noop {} {\bibfield  {journal} {\bibinfo  {journal} {Nanoscale}\
  }\textbf {\bibinfo {volume} {10}},\ \bibinfo {pages} {13565} (\bibinfo {year}
  {2018})}\BibitemShut {NoStop}%
\bibitem [{\citenamefont {Xing}\ \emph {et~al.}(2018)\citenamefont {Xing},
  \citenamefont {Zhao}, \citenamefont {Askerka}, \citenamefont {Quan},
  \citenamefont {Gong}, \citenamefont {Zhao}, \citenamefont {Zhao},
  \citenamefont {Tan}, \citenamefont {Long}, \citenamefont {Gao} \emph
  {et~al.}}]{xing2018color}%
  \BibitemOpen
  \bibfield  {author} {\bibinfo {author} {\bibfnamefont {J.}~\bibnamefont
  {Xing}}, \bibinfo {author} {\bibfnamefont {Y.}~\bibnamefont {Zhao}}, \bibinfo
  {author} {\bibfnamefont {M.}~\bibnamefont {Askerka}}, \bibinfo {author}
  {\bibfnamefont {L.~N.}\ \bibnamefont {Quan}}, \bibinfo {author}
  {\bibfnamefont {X.}~\bibnamefont {Gong}}, \bibinfo {author} {\bibfnamefont
  {W.}~\bibnamefont {Zhao}}, \bibinfo {author} {\bibfnamefont {J.}~\bibnamefont
  {Zhao}}, \bibinfo {author} {\bibfnamefont {H.}~\bibnamefont {Tan}}, \bibinfo
  {author} {\bibfnamefont {G.}~\bibnamefont {Long}}, \bibinfo {author}
  {\bibfnamefont {L.}~\bibnamefont {Gao}}, \emph {et~al.},\ }\bibfield  {title}
  {\bibinfo {title} {Color-stable highly luminescent sky-blue perovskite
  light-emitting diodes},\ }\href@noop {} {\bibfield  {journal} {\bibinfo
  {journal} {Nature communications}\ }\textbf {\bibinfo {volume} {9}},\
  \bibinfo {pages} {3541} (\bibinfo {year} {2018})}\BibitemShut {NoStop}%
\bibitem [{\citenamefont {Chakraborty}\ \emph {et~al.}(2022)\citenamefont
  {Chakraborty}, \citenamefont {Paul},\ and\ \citenamefont
  {Pal}}]{PhysRevApplied.17.054045}%
  \BibitemOpen
  \bibfield  {author} {\bibinfo {author} {\bibfnamefont {R.}~\bibnamefont
  {Chakraborty}}, \bibinfo {author} {\bibfnamefont {G.}~\bibnamefont {Paul}},\
  and\ \bibinfo {author} {\bibfnamefont {A.~J.}\ \bibnamefont {Pal}},\
  }\bibfield  {title} {\bibinfo {title} {Quantum confinement and dielectric
  deconfinement in quasi-two-dimensional perovskites: Their roles in
  light-emitting diodes},\ }\href
  {https://doi.org/10.1103/PhysRevApplied.17.054045} {\bibfield  {journal}
  {\bibinfo  {journal} {Phys. Rev. Appl.}\ }\textbf {\bibinfo {volume} {17}},\
  \bibinfo {pages} {054045} (\bibinfo {year} {2022})}\BibitemShut {NoStop}%
\bibitem [{\citenamefont {Nabi}\ \emph {et~al.}(2024)\citenamefont {Nabi},
  \citenamefont {Padelkar}, \citenamefont {Jasieniak}, \citenamefont
  {Simonov},\ and\ \citenamefont {Alam}}]{PhysRevApplied.21.014063}%
  \BibitemOpen
  \bibfield  {author} {\bibinfo {author} {\bibfnamefont {M.}~\bibnamefont
  {Nabi}}, \bibinfo {author} {\bibfnamefont {S.~S.}\ \bibnamefont {Padelkar}},
  \bibinfo {author} {\bibfnamefont {J.~J.}\ \bibnamefont {Jasieniak}}, \bibinfo
  {author} {\bibfnamefont {A.~N.}\ \bibnamefont {Simonov}},\ and\ \bibinfo
  {author} {\bibfnamefont {A.}~\bibnamefont {Alam}},\ }\bibfield  {title}
  {\bibinfo {title} {Lead-free magnetic double perovskites for photovoltaic and
  photocatalysis applications},\ }\href
  {https://doi.org/10.1103/PhysRevApplied.21.014063} {\bibfield  {journal}
  {\bibinfo  {journal} {Phys. Rev. Appl.}\ }\textbf {\bibinfo {volume} {21}},\
  \bibinfo {pages} {014063} (\bibinfo {year} {2024})}\BibitemShut {NoStop}%
\bibitem [{\citenamefont {Saidaminov}\ \emph {et~al.}(2015)\citenamefont
  {Saidaminov}, \citenamefont {Adinolfi}, \citenamefont {Comin}, \citenamefont
  {Abdelhady}, \citenamefont {Peng}, \citenamefont {Dursun}, \citenamefont
  {Yuan}, \citenamefont {Hoogland}, \citenamefont {Sargent},\ and\
  \citenamefont {Bakr}}]{saidaminov2015planar}%
  \BibitemOpen
  \bibfield  {author} {\bibinfo {author} {\bibfnamefont {M.~I.}\ \bibnamefont
  {Saidaminov}}, \bibinfo {author} {\bibfnamefont {V.}~\bibnamefont
  {Adinolfi}}, \bibinfo {author} {\bibfnamefont {R.}~\bibnamefont {Comin}},
  \bibinfo {author} {\bibfnamefont {A.~L.}\ \bibnamefont {Abdelhady}}, \bibinfo
  {author} {\bibfnamefont {W.}~\bibnamefont {Peng}}, \bibinfo {author}
  {\bibfnamefont {I.}~\bibnamefont {Dursun}}, \bibinfo {author} {\bibfnamefont
  {M.}~\bibnamefont {Yuan}}, \bibinfo {author} {\bibfnamefont {S.}~\bibnamefont
  {Hoogland}}, \bibinfo {author} {\bibfnamefont {E.~H.}\ \bibnamefont
  {Sargent}},\ and\ \bibinfo {author} {\bibfnamefont {O.~M.}\ \bibnamefont
  {Bakr}},\ }\bibfield  {title} {\bibinfo {title} {Planar-integrated
  single-crystalline perovskite photodetectors},\ }\href@noop {} {\bibfield
  {journal} {\bibinfo  {journal} {Nature communications}\ }\textbf {\bibinfo
  {volume} {6}},\ \bibinfo {pages} {8724} (\bibinfo {year} {2015})}\BibitemShut
  {NoStop}%
\bibitem [{\citenamefont {Leijtens}\ \emph {et~al.}(2015)\citenamefont
  {Leijtens}, \citenamefont {Eperon}, \citenamefont {Noel}, \citenamefont
  {Habisreutinger}, \citenamefont {Petrozza},\ and\ \citenamefont
  {Snaith}}]{leijtens2015stability}%
  \BibitemOpen
  \bibfield  {author} {\bibinfo {author} {\bibfnamefont {T.}~\bibnamefont
  {Leijtens}}, \bibinfo {author} {\bibfnamefont {G.~E.}\ \bibnamefont
  {Eperon}}, \bibinfo {author} {\bibfnamefont {N.~K.}\ \bibnamefont {Noel}},
  \bibinfo {author} {\bibfnamefont {S.~N.}\ \bibnamefont {Habisreutinger}},
  \bibinfo {author} {\bibfnamefont {A.}~\bibnamefont {Petrozza}},\ and\
  \bibinfo {author} {\bibfnamefont {H.~J.}\ \bibnamefont {Snaith}},\ }\bibfield
   {title} {\bibinfo {title} {Stability of metal halide perovskite solar
  cells},\ }\href@noop {} {\bibfield  {journal} {\bibinfo  {journal} {Advanced
  Energy Materials}\ }\textbf {\bibinfo {volume} {5}},\ \bibinfo {pages}
  {1500963} (\bibinfo {year} {2015})}\BibitemShut {NoStop}%
\bibitem [{\citenamefont {Berhe}\ \emph {et~al.}(2016)\citenamefont {Berhe},
  \citenamefont {Su}, \citenamefont {Chen}, \citenamefont {Pan}, \citenamefont
  {Cheng}, \citenamefont {Chen}, \citenamefont {Tsai}, \citenamefont {Chen},
  \citenamefont {Dubale},\ and\ \citenamefont {Hwang}}]{berhe2016organometal}%
  \BibitemOpen
  \bibfield  {author} {\bibinfo {author} {\bibfnamefont {T.~A.}\ \bibnamefont
  {Berhe}}, \bibinfo {author} {\bibfnamefont {W.-N.}\ \bibnamefont {Su}},
  \bibinfo {author} {\bibfnamefont {C.-H.}\ \bibnamefont {Chen}}, \bibinfo
  {author} {\bibfnamefont {C.-J.}\ \bibnamefont {Pan}}, \bibinfo {author}
  {\bibfnamefont {J.-H.}\ \bibnamefont {Cheng}}, \bibinfo {author}
  {\bibfnamefont {H.-M.}\ \bibnamefont {Chen}}, \bibinfo {author}
  {\bibfnamefont {M.-C.}\ \bibnamefont {Tsai}}, \bibinfo {author}
  {\bibfnamefont {L.-Y.}\ \bibnamefont {Chen}}, \bibinfo {author}
  {\bibfnamefont {A.~A.}\ \bibnamefont {Dubale}},\ and\ \bibinfo {author}
  {\bibfnamefont {B.-J.}\ \bibnamefont {Hwang}},\ }\bibfield  {title} {\bibinfo
  {title} {Organometal halide perovskite solar cells: degradation and
  stability},\ }\href@noop {} {\bibfield  {journal} {\bibinfo  {journal}
  {Energy \& Environmental Science}\ }\textbf {\bibinfo {volume} {9}},\
  \bibinfo {pages} {323} (\bibinfo {year} {2016})}\BibitemShut {NoStop}%
\bibitem [{\citenamefont {Wang}\ \emph {et~al.}(2016)\citenamefont {Wang},
  \citenamefont {Wright}, \citenamefont {Elumalai},\ and\ \citenamefont
  {Uddin}}]{wang2016stability}%
  \BibitemOpen
  \bibfield  {author} {\bibinfo {author} {\bibfnamefont {D.}~\bibnamefont
  {Wang}}, \bibinfo {author} {\bibfnamefont {M.}~\bibnamefont {Wright}},
  \bibinfo {author} {\bibfnamefont {N.~K.}\ \bibnamefont {Elumalai}},\ and\
  \bibinfo {author} {\bibfnamefont {A.}~\bibnamefont {Uddin}},\ }\bibfield
  {title} {\bibinfo {title} {Stability of perovskite solar cells},\ }\href@noop
  {} {\bibfield  {journal} {\bibinfo  {journal} {Solar Energy Materials and
  Solar Cells}\ }\textbf {\bibinfo {volume} {147}},\ \bibinfo {pages} {255}
  (\bibinfo {year} {2016})}\BibitemShut {NoStop}%
\bibitem [{\citenamefont {Guarnera}\ \emph {et~al.}(2015)\citenamefont
  {Guarnera}, \citenamefont {Abate}, \citenamefont {Zhang}, \citenamefont
  {Foster}, \citenamefont {Richardson}, \citenamefont {Petrozza},\ and\
  \citenamefont {Snaith}}]{guarnera2015improving}%
  \BibitemOpen
  \bibfield  {author} {\bibinfo {author} {\bibfnamefont {S.}~\bibnamefont
  {Guarnera}}, \bibinfo {author} {\bibfnamefont {A.}~\bibnamefont {Abate}},
  \bibinfo {author} {\bibfnamefont {W.}~\bibnamefont {Zhang}}, \bibinfo
  {author} {\bibfnamefont {J.~M.}\ \bibnamefont {Foster}}, \bibinfo {author}
  {\bibfnamefont {G.}~\bibnamefont {Richardson}}, \bibinfo {author}
  {\bibfnamefont {A.}~\bibnamefont {Petrozza}},\ and\ \bibinfo {author}
  {\bibfnamefont {H.~J.}\ \bibnamefont {Snaith}},\ }\bibfield  {title}
  {\bibinfo {title} {Improving the long-term stability of perovskite solar
  cells with a porous al2o3 buffer layer},\ }\href@noop {} {\bibfield
  {journal} {\bibinfo  {journal} {The journal of physical chemistry letters}\
  }\textbf {\bibinfo {volume} {6}},\ \bibinfo {pages} {432} (\bibinfo {year}
  {2015})}\BibitemShut {NoStop}%
\bibitem [{\citenamefont {Huang}\ \emph {et~al.}(2016)\citenamefont {Huang},
  \citenamefont {Li}, \citenamefont {Kong}, \citenamefont {Zhu}, \citenamefont
  {Shan},\ and\ \citenamefont {Li}}]{huang2016enhancing}%
  \BibitemOpen
  \bibfield  {author} {\bibinfo {author} {\bibfnamefont {S.}~\bibnamefont
  {Huang}}, \bibinfo {author} {\bibfnamefont {Z.}~\bibnamefont {Li}}, \bibinfo
  {author} {\bibfnamefont {L.}~\bibnamefont {Kong}}, \bibinfo {author}
  {\bibfnamefont {N.}~\bibnamefont {Zhu}}, \bibinfo {author} {\bibfnamefont
  {A.}~\bibnamefont {Shan}},\ and\ \bibinfo {author} {\bibfnamefont
  {L.}~\bibnamefont {Li}},\ }\bibfield  {title} {\bibinfo {title} {Enhancing
  the stability of {CH}$_3${NH}$_3${P}b{B}r$_3$ quantum dots by embedding in
  silica spheres derived from tetramethyl orthosilicate in “waterless”
  toluene},\ }\href@noop {} {\bibfield  {journal} {\bibinfo  {journal} {Journal
  of the American Chemical Society}\ }\textbf {\bibinfo {volume} {138}},\
  \bibinfo {pages} {5749} (\bibinfo {year} {2016})}\BibitemShut {NoStop}%
\bibitem [{\citenamefont {Di}\ \emph {et~al.}(2015)\citenamefont {Di},
  \citenamefont {Musselman}, \citenamefont {Li}, \citenamefont {Sadhanala},
  \citenamefont {Ievskaya}, \citenamefont {Song}, \citenamefont {Tan},
  \citenamefont {Lai}, \citenamefont {MacManus-Driscoll}, \citenamefont
  {Greenham} \emph {et~al.}}]{di2015size}%
  \BibitemOpen
  \bibfield  {author} {\bibinfo {author} {\bibfnamefont {D.}~\bibnamefont
  {Di}}, \bibinfo {author} {\bibfnamefont {K.~P.}\ \bibnamefont {Musselman}},
  \bibinfo {author} {\bibfnamefont {G.}~\bibnamefont {Li}}, \bibinfo {author}
  {\bibfnamefont {A.}~\bibnamefont {Sadhanala}}, \bibinfo {author}
  {\bibfnamefont {Y.}~\bibnamefont {Ievskaya}}, \bibinfo {author}
  {\bibfnamefont {Q.}~\bibnamefont {Song}}, \bibinfo {author} {\bibfnamefont
  {Z.-K.}\ \bibnamefont {Tan}}, \bibinfo {author} {\bibfnamefont {M.~L.}\
  \bibnamefont {Lai}}, \bibinfo {author} {\bibfnamefont {J.~L.}\ \bibnamefont
  {MacManus-Driscoll}}, \bibinfo {author} {\bibfnamefont {N.~C.}\ \bibnamefont
  {Greenham}}, \emph {et~al.},\ }\bibfield  {title} {\bibinfo {title}
  {Size-dependent photon emission from organometal halide perovskite
  nanocrystals embedded in an organic matrix},\ }\href@noop {} {\bibfield
  {journal} {\bibinfo  {journal} {The journal of physical chemistry letters}\
  }\textbf {\bibinfo {volume} {6}},\ \bibinfo {pages} {446} (\bibinfo {year}
  {2015})}\BibitemShut {NoStop}%
\bibitem [{\citenamefont {Tannenbaum}\ \emph {et~al.}(2006)\citenamefont
  {Tannenbaum}, \citenamefont {Zubris}, \citenamefont {David}, \citenamefont
  {Ciprari}, \citenamefont {Jacob}, \citenamefont {Jasiuk},\ and\ \citenamefont
  {Dan}}]{tannenbaum2006ftir}%
  \BibitemOpen
  \bibfield  {author} {\bibinfo {author} {\bibfnamefont {R.}~\bibnamefont
  {Tannenbaum}}, \bibinfo {author} {\bibfnamefont {M.}~\bibnamefont {Zubris}},
  \bibinfo {author} {\bibfnamefont {K.}~\bibnamefont {David}}, \bibinfo
  {author} {\bibfnamefont {D.}~\bibnamefont {Ciprari}}, \bibinfo {author}
  {\bibfnamefont {K.}~\bibnamefont {Jacob}}, \bibinfo {author} {\bibfnamefont
  {I.}~\bibnamefont {Jasiuk}},\ and\ \bibinfo {author} {\bibfnamefont
  {N.}~\bibnamefont {Dan}},\ }\bibfield  {title} {\bibinfo {title} {Ftir
  characterization of the reactive interface of cobalt oxide nanoparticles
  embedded in polymeric matrices},\ }\href@noop {} {\bibfield  {journal}
  {\bibinfo  {journal} {The Journal of Physical Chemistry B}\ }\textbf
  {\bibinfo {volume} {110}},\ \bibinfo {pages} {2227} (\bibinfo {year}
  {2006})}\BibitemShut {NoStop}%
\bibitem [{\citenamefont {Tannenbaum}\ \emph {et~al.}(2004)\citenamefont
  {Tannenbaum}, \citenamefont {King}, \citenamefont {Lecy}, \citenamefont
  {Tirrell},\ and\ \citenamefont {Potts}}]{tannenbaum2004infrared}%
  \BibitemOpen
  \bibfield  {author} {\bibinfo {author} {\bibfnamefont {R.}~\bibnamefont
  {Tannenbaum}}, \bibinfo {author} {\bibfnamefont {S.}~\bibnamefont {King}},
  \bibinfo {author} {\bibfnamefont {J.}~\bibnamefont {Lecy}}, \bibinfo {author}
  {\bibfnamefont {M.}~\bibnamefont {Tirrell}},\ and\ \bibinfo {author}
  {\bibfnamefont {L.}~\bibnamefont {Potts}},\ }\bibfield  {title} {\bibinfo
  {title} {Infrared study of the kinetics and mechanism of adsorption of
  acrylic polymers on alumina surfaces},\ }\href@noop {} {\bibfield  {journal}
  {\bibinfo  {journal} {Langmuir}\ }\textbf {\bibinfo {volume} {20}},\ \bibinfo
  {pages} {4507} (\bibinfo {year} {2004})}\BibitemShut {NoStop}%
\bibitem [{\citenamefont {Ciprari}\ \emph {et~al.}(2006)\citenamefont
  {Ciprari}, \citenamefont {Jacob},\ and\ \citenamefont
  {Tannenbaum}}]{ciprari2006characterization}%
  \BibitemOpen
  \bibfield  {author} {\bibinfo {author} {\bibfnamefont {D.}~\bibnamefont
  {Ciprari}}, \bibinfo {author} {\bibfnamefont {K.}~\bibnamefont {Jacob}},\
  and\ \bibinfo {author} {\bibfnamefont {R.}~\bibnamefont {Tannenbaum}},\
  }\bibfield  {title} {\bibinfo {title} {Characterization of polymer
  nanocomposite interphase and its impact on mechanical properties},\
  }\href@noop {} {\bibfield  {journal} {\bibinfo  {journal} {Macromolecules}\
  }\textbf {\bibinfo {volume} {39}},\ \bibinfo {pages} {6565} (\bibinfo {year}
  {2006})}\BibitemShut {NoStop}%
\bibitem [{\citenamefont {Li}\ \emph {et~al.}(2018{\natexlab{b}})\citenamefont
  {Li}, \citenamefont {Xue}, \citenamefont {Luo}, \citenamefont {Huang},
  \citenamefont {Liu}, \citenamefont {Qiao}, \citenamefont {Liu}, \citenamefont
  {Song}, \citenamefont {Yan}, \citenamefont {Li} \emph
  {et~al.}}]{li2018stable}%
  \BibitemOpen
  \bibfield  {author} {\bibinfo {author} {\bibfnamefont {X.}~\bibnamefont
  {Li}}, \bibinfo {author} {\bibfnamefont {Z.}~\bibnamefont {Xue}}, \bibinfo
  {author} {\bibfnamefont {D.}~\bibnamefont {Luo}}, \bibinfo {author}
  {\bibfnamefont {C.}~\bibnamefont {Huang}}, \bibinfo {author} {\bibfnamefont
  {L.}~\bibnamefont {Liu}}, \bibinfo {author} {\bibfnamefont {X.}~\bibnamefont
  {Qiao}}, \bibinfo {author} {\bibfnamefont {C.}~\bibnamefont {Liu}}, \bibinfo
  {author} {\bibfnamefont {Q.}~\bibnamefont {Song}}, \bibinfo {author}
  {\bibfnamefont {C.}~\bibnamefont {Yan}}, \bibinfo {author} {\bibfnamefont
  {Y.}~\bibnamefont {Li}}, \emph {et~al.},\ }\bibfield  {title} {\bibinfo
  {title} {A stable lead halide perovskite nanocrystals protected by pmma},\
  }\href@noop {} {\bibfield  {journal} {\bibinfo  {journal} {Sci. China Mater}\
  }\textbf {\bibinfo {volume} {61}},\ \bibinfo {pages} {363} (\bibinfo {year}
  {2018}{\natexlab{b}})}\BibitemShut {NoStop}%
\bibitem [{\citenamefont {Wei}\ \emph {et~al.}(2017)\citenamefont {Wei},
  \citenamefont {Tang}, \citenamefont {Feng},\ and\ \citenamefont
  {You}}]{wei2017importance}%
  \BibitemOpen
  \bibfield  {author} {\bibinfo {author} {\bibfnamefont {H.}~\bibnamefont
  {Wei}}, \bibinfo {author} {\bibfnamefont {Y.}~\bibnamefont {Tang}}, \bibinfo
  {author} {\bibfnamefont {B.}~\bibnamefont {Feng}},\ and\ \bibinfo {author}
  {\bibfnamefont {H.}~\bibnamefont {You}},\ }\bibfield  {title} {\bibinfo
  {title} {Importance of pbi2 morphology in two-step deposition of ch3nh3pbi3
  for high-performance perovskite solar cells},\ }\href@noop {} {\bibfield
  {journal} {\bibinfo  {journal} {Chinese Physics B}\ }\textbf {\bibinfo
  {volume} {26}},\ \bibinfo {pages} {128801} (\bibinfo {year}
  {2017})}\BibitemShut {NoStop}%
\end{thebibliography}%

\end{document}